\documentclass[aps,floatfix,twocolumn]{revtex4-1}

\usepackage[utf8]{inputenc}
\usepackage{amsmath,amssymb} 
\usepackage{hyperref}
\usepackage{graphicx}
\usepackage{float}
\usepackage{appendix}

\DeclareUnicodeCharacter{03B4}{$\Delta$}
\DeclareUnicodeCharacter{2212}{-}

\begin{document}

\title{Localization, transport and edge states in a two-strand ladder network in an aperiodically staggered magnetic field}
 
\author{Sk Sajid}
\email{sajid.burd@gmail.com}
\author{Arunava Chakrabarti}
\email{arunava.physics@presiuniv.ac.in}
\affiliation{ Department of Physics, Presidency University, 86/1 College Street, Kolkata-700073, INDIA}

\date{\today}

\begin{abstract}
We investigate the spectral and transport properties of a two-arm tight-binding ladder perturbed by an external magnetic field following an Aubry-Andr\'e-Harper profile. The varying magnetic flux trapped in consecutive ladder-cells simulates an axial twist that enables us, in principle, to probe a wide variety of systems ranging from a ribbon Hofstadter geometry to helical DNA chains. We perform an in-depth numerical analysis, using a direct diagonalization of the lattice Hamiltonian to study the electronic spectra and transport properties of the model. We show that such a geometry creates a self-similar multifractal pattern in the energy landscape. The spectral properties are analyzed using the local density of states and a Green’s function formalism is employed to obtain the two-terminal transmission probability. With the standard multifractal analysis and the evaluation of inverse participation ratio we show that, the system hosts both critical and extended phase for a slowly varying aperiodic sequence of flux indicating a possible mobility edge. Finally, we report signatures of topological edge modes that are found to be robust against a correlated perturbation given to the nearest neighbor hopping integrals. Our results can be of importance in experiments involving ladder-like quantum networks, realized with cold atoms in an optical trap setup.
\end{abstract}

\maketitle



\section{Introduction}
	The discovery of quantum Hall effect~\cite{klitzing} has unleashed a plethora of theoretical investigations and experimental observations of different non-trivial topological phases ~\cite{bernevig,konig,kane,hsieh} making it one of the most active research domains in the present day condensed matter physics. Such systems host non-trivial quantum phases, characterized by topological edge states on their surface, that are robust against any local perturbation. This remarkable fact makes such systems potential materials where a transportation of energy can be engineered with controlled or mitigated dissipation.\\
 One of the most celebrated models in the context of topological phases was given by the pioneering work of Hofstadter~\cite{hofstadter}. His work describes two-dimensional electron gas in a periodic lattice potential. When subjected to a strong magnetic field, the interplay of two length-scales causes the energy bands to fragment into smaller subbands producing a self-similar fractal pattern, famously known as the 'Hofstadter butterfly'. However, the butterfly pattern is only observed if the magnetic length scale is much smaller than the underlying lattice periodicity requiring fluxes $\phi$ to be a significant fraction of $\phi_o$. For conventional condensed matter lattices, this translates to an astronomically large magnetic field which prevented researchers to realize the butterfly experimentally for decades. An alternative approach involves the use of superlattice structures that effectively increases the lattice constant to lower the requirement of magnetic field. This key concept has made it possible to realize the butterfly in graphene~\cite{dean,hunt} and semiconductor superlattice~\cite{schlosser} structures. Even, reconfigurable quasi-periodic acoustic crystals have recently been realized and are reported to show Hofstadter butterfly and topological edge states, taking the excitement beyond the realm of conventional electronic systems~\cite{xiang}\\
 Recently, a lower dimensional version of this problem, viz, the Hofstadter model on a strip, has gained a lot of interest and, has triggered a series of theoretical~\cite{lau,mugel} and experimental~\cite{genika,tai} investigation focusing on the fate of the Hofstadter butterfly and the nature of transport when the motion of the electron is severely restricted in one or more directions. However, previous works on strip geometries  mostly concentrate on a flat ribbon-like structure that lack generality in terms of flux distribution and completely neglects any effect that the structural geometry may have on this problem. The effect of geometry in the context of low-dimensional physics can be a useful consideration while modeling biological systems such as a DNA molecule.\\
The question of whether a DNA conducts electric charge has kindled the interest of Physicists and Biologists over the years~\cite{berlin,delaney,endres} and remains a challenging task to address even now. An understanding of the charge migration process may provide important insight about the biological processes such as DNA damage repair~\cite{boon}. However, modelling an actual DNA strand is a daunting task given its extremely complicated structure. One of the ways to attack such problems is to make a suitable model of the system first and then use the tight-binding (TB) approximation~\cite{romer2,romer3}. As a first approximation, one can treat each Watson-crick base pair a an effective, single site. This enables one to incorporate the basic features of a DNA in a tight binding Hamiltonian~\cite{macia} where a suitable value of the on-site potential is attributed to the lattice site (here, the sites in the ladder), along with an appropriate hopping amplitude describing the overlap of the Wannier orbitals between the neighboring sites. The hydrogen bonding between the complementary bases can then be mimicked by the inter-strand hopping of the ladder network. However, apart from modeling DNA or proteins the  tight-binding description has been used in numerous theoretical studies to mimic a wide variety of systems such as non-linear waveguides~\cite{kivshar}, Bose-Einstein condensate in optical lattices~\cite{andrea}, and, superconducting Josephson junctions~\cite{binder}.

In general, ladder networks have a uni-directional geometry with only a few sites attached in the perpendicular direction. The resulting sharp boundaries make such a network an excellent candidate for detection of the topological edge features~\cite{stuhl}. What is more important is that, these structures are not just mere theoretical tools but can be readily generated in laboratories with the coherent and sequential coupling of internal states of atoms.\\
This backdrop motivates us to examine the transport and topological edge state features of a two-arm tight-binding ladder network trapping a magnetic flux in each cell, that has a staggered distribution. We have chosen a generalization of the Aubry-Andr\'{e}-Harper (AAH) modulation in the flux distribution, that encodes an effective {\it tunable axial twist} in the ladder network. Originally, the AAH model was proposed as a paradigmatic model of an incommensurate one dimensional system, that exhibits a metal-insulator transition in parameter space~\cite{aubry,harper}. In recent times, the AAH potential profile, and its variants are also being thought of as suitable for developing optical lattices~\cite{modungo}, in describing bichromatic potentials~\cite{patricio}, and in engineered parity-time symmetric photonic lattices~\cite{regensburger}, to cite a few examples.

In this paper, we show that even such a simple construction of modulated flux mapped on a quasi-1D Hofstadter strip gives rise to plenty of rich physics. The real-space diagonalization of the Hamiltonian reveals that the Hofstadter butterfly survives in its lower-dimensional counterpart. To understand the role of geometry on the spectral and transport properties, we estimate the local density of states (LDOS) and multi-terminal transport for different distributions of flux.  Our quantum transport calculation and inverse participation ratio (IPR) analysis reveal discrete regions of extended and critical phases in the energy spectrum and thus suggests the existence of mobility edges. We also perform a multifractal analysis to further support our claims of quantum phases. Since the original AAH model has an inversion symmetry in its commensurate limit, our proposed system guarantees topological edge doublets that are chiral in nature. To test the robustness, we introduce a staggered perturbation in the hopping integrals that simulates a physical deformation of the ladder geometry. Finally, we dedicate a section in the end on how to realize our work experimentally.

\section{The model}
\begin{figure}[h!]
	\includegraphics[width=\columnwidth]{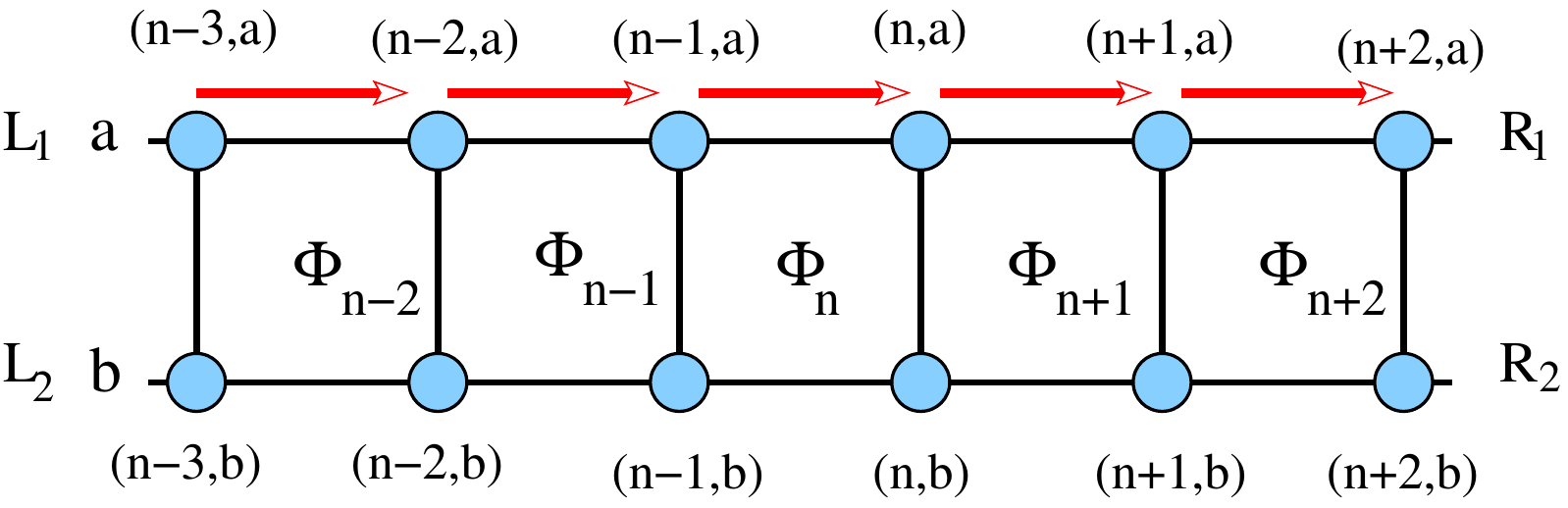}
	\vspace{-0.5cm}
	\caption{(Color online.) Schematic diagram of a non-interacting spinless tight-binding ladder model in a perpendicular magnetic field. The flux at the nth plaquette is $\Phi_n$ and only affects the electrons moving along the a-leg due to gauge choice. The lead-connections are marked as $L_i$ and $R_i$ ($i=1$,$2$), and are discussed in the text.} 
	\label{fig:ladder}
\end{figure}

We consider a system of non-interacting, spinless particles in a two-arm ladder network, subjected to a magnetic field perpendicular to the ladder plane [see Fig-\ref{fig:ladder}]. The ladder has N plaquettes or $n= 2(N+1)$ atomic sites ($i=1,2,3...n$) and assumes open boundary conditions primarily.

 To incorporate the angular twist, we introduce an AAH modulation profile in the magnetic flux trapped in the ladder cells. For convenience, and definitely without losing any physics, we choose to write the flux per plaquette, normalized by the flux quantum $\Phi_0=hc/e$, as 
\begin{equation}\label{eq:flux}
	\Phi_i=(\lambda/2\pi) \cos(2 \pi \alpha i^{\nu})
\end{equation}
Here $\lambda$ is the `strength' of the modulation, and  $0\leq\nu\leq 1$ defines a `slowness parameter' $\nu$. The twist parameter $\alpha$ controls the frequency of the flux modulation. 
\begin{figure*}[htpb]
	\centering
	(a)\includegraphics[width=0.6\columnwidth]{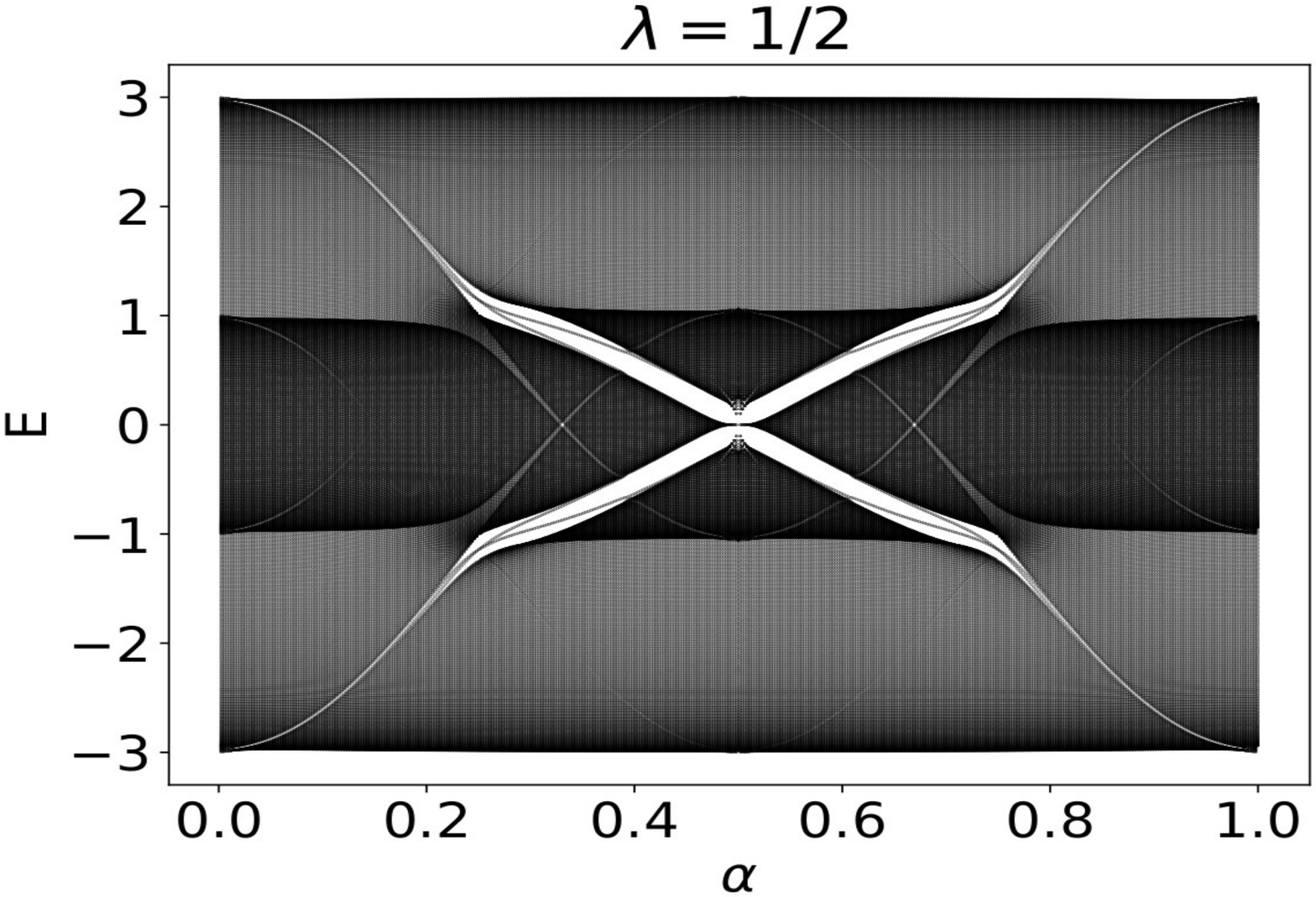}
	(b)\includegraphics[width=0.6\columnwidth]{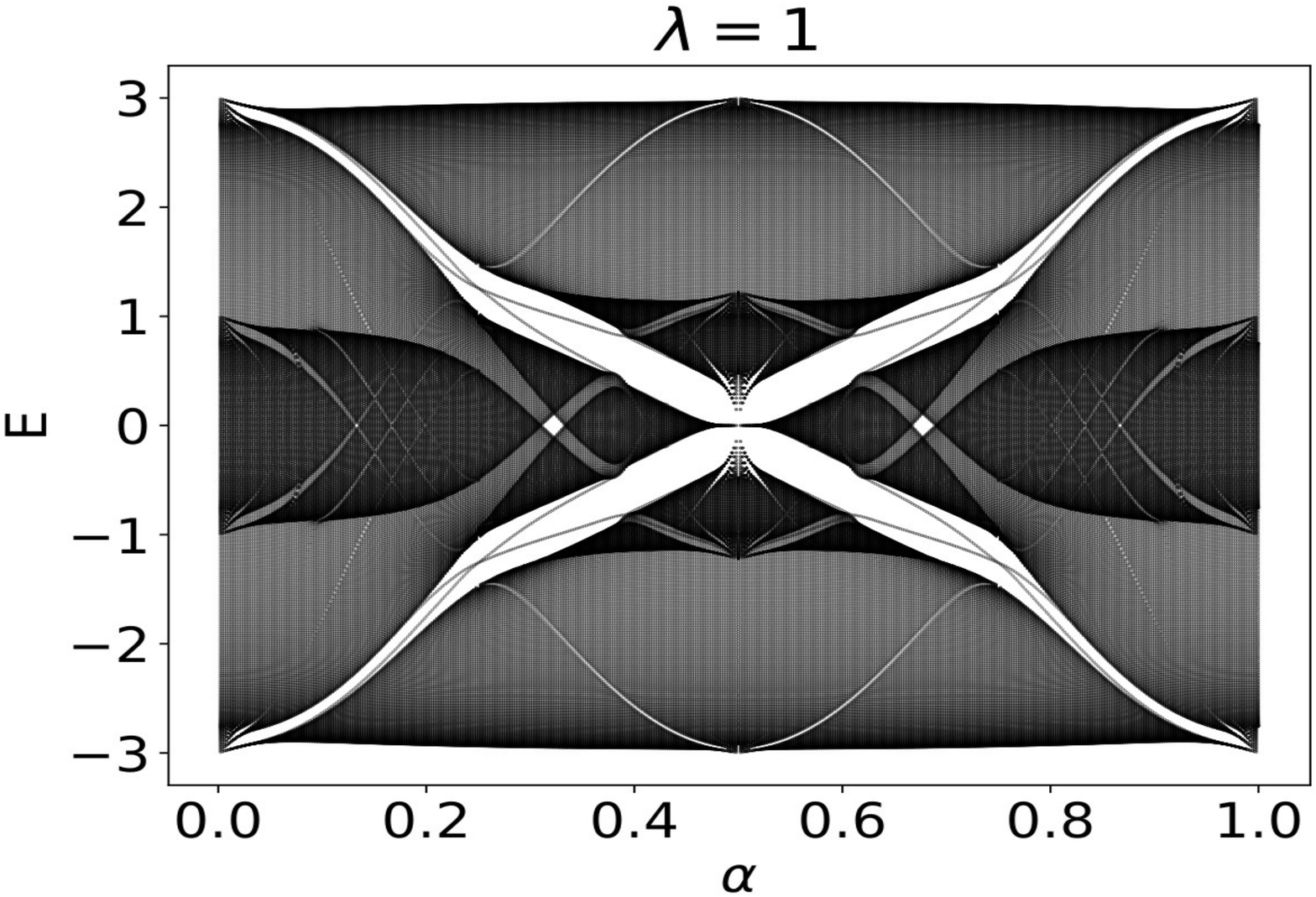}
	\\
	(c)\includegraphics[width=0.6\columnwidth]{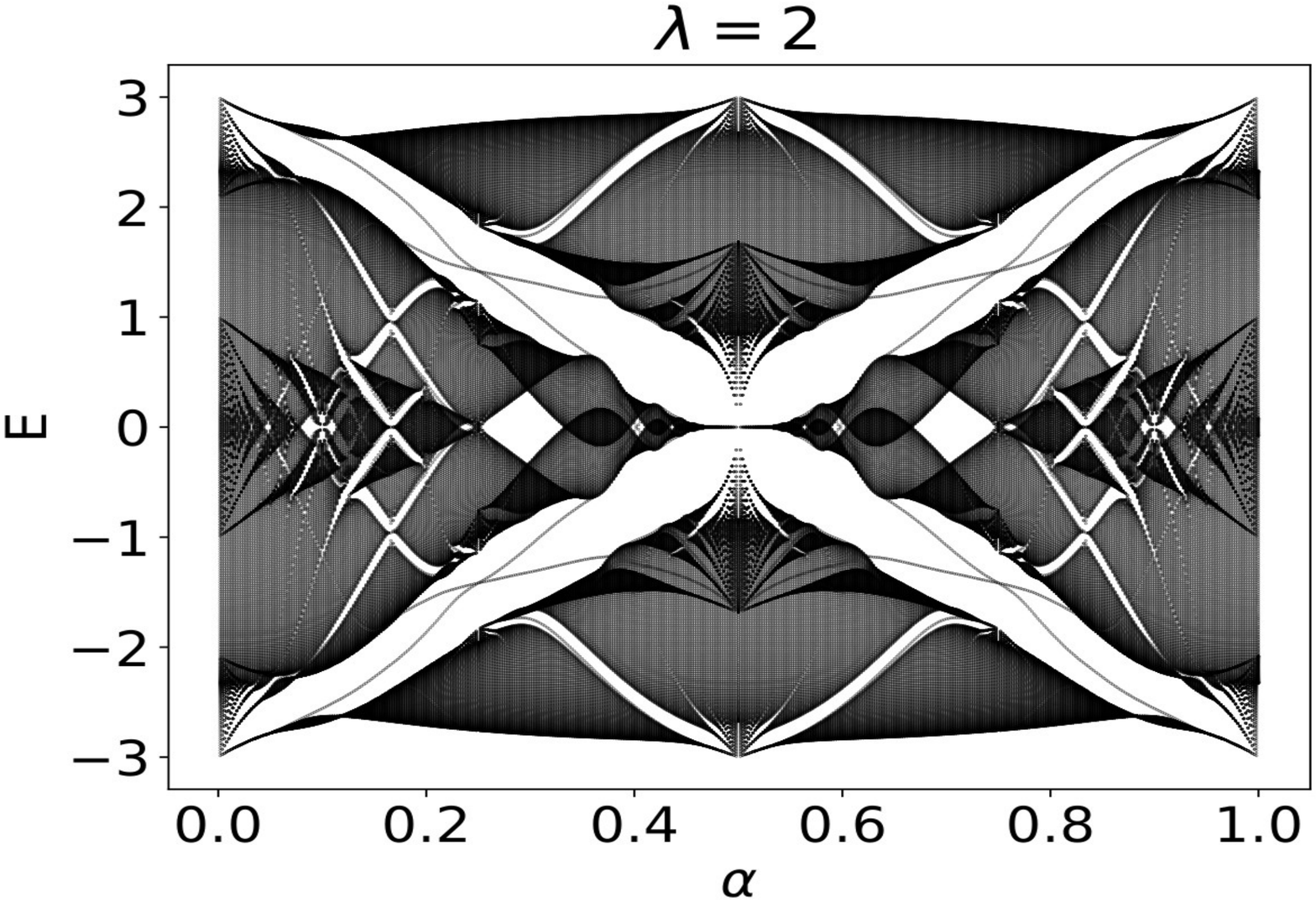}
	(d)\includegraphics[width=0.6\columnwidth]{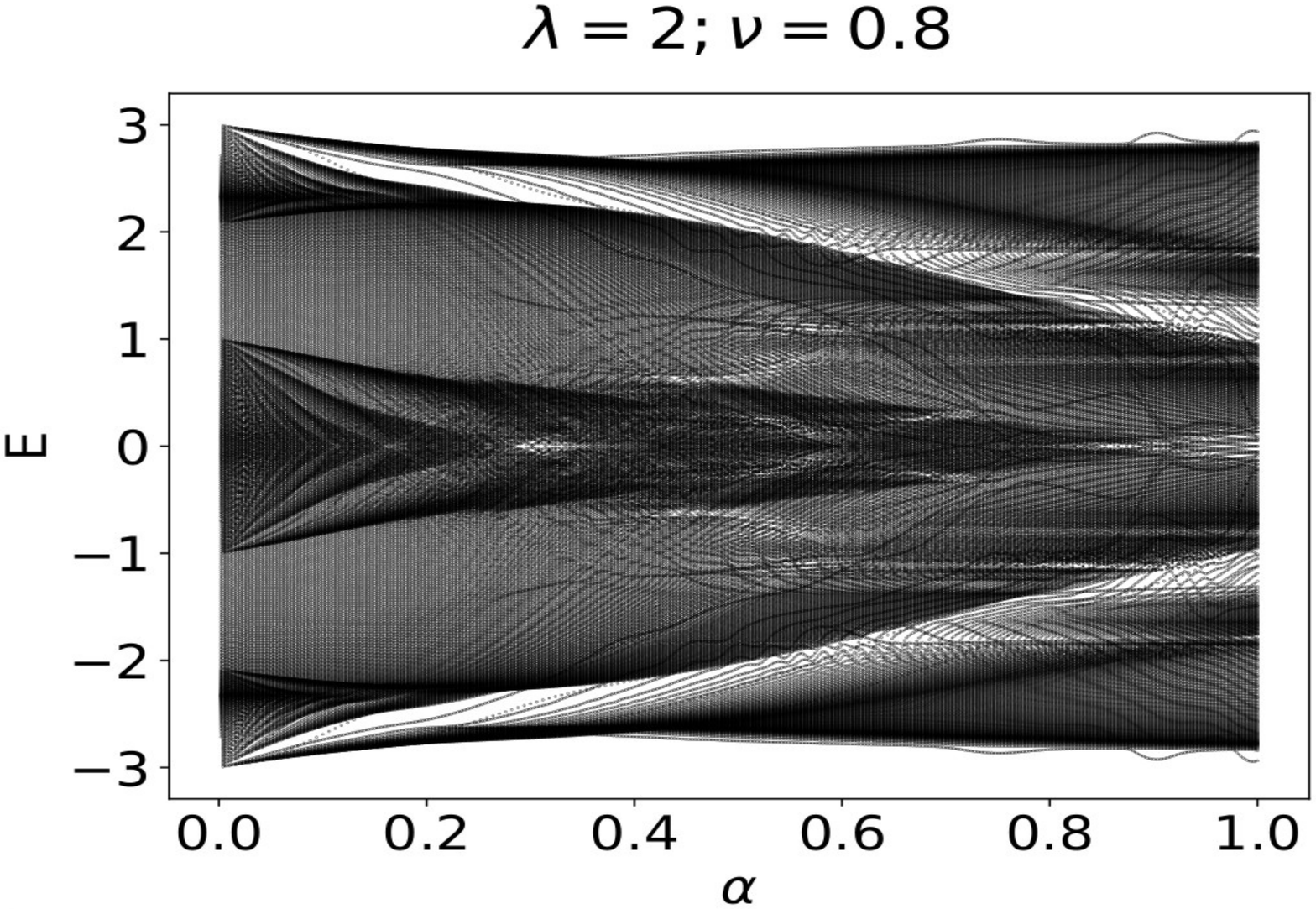}
	\caption{Single-particle energy spectra E vs flux modulation frequency $\alpha$ are plotted for various $\lambda$. (a) The spectrum is overall continuous with a small band-gap for $\lambda=1/2$, and $\nu=1$, (b) the band-gap gets broadened if $\lambda=1$, keeping $\nu=1$ , and (c) at $\lambda=2$, with $\nu=1$ still, the band-gap starts to fragment into smaller subbands exhibiting fractal nature. (d) For a slower flux modulation, with  $\nu=0.8$,  the interference pattern is destroyed.} 
	\label{fig:butterfly}
\end{figure*}
The limit $\nu = 0$ represents a constant flux distribution for a given $\alpha$, and the other limit of $\nu=1$ brings back the AAH modulation in the same. It is worth mentioning that, a similar slowly varying function (SVF) was first introduced in designing an aperiodic potential landscape~\cite{dassarma2} in a one dimensional tight binding chain of atomic sites. It was argued that, a non-zero value of $\nu$ triggered a metal-insulator transition in a one dimensional chain of atoms, and one could work out the existence of mobility edges~\cite{dassarma}, a result that is not obtained in the conventional AAH model. To the best of our knowledge, the effect of such a modulation in the distribution of the magnetic flux trapped in a two-strand ladder has not been addressed before.

This choice of correlation in the flux also serves as a general function for the realization of different flux distributions in the plaquettes. For instance, with $\nu=1$, one may set $\alpha$ to 1, 1/2 or any irrational number to obtain uniform, staggered or quasiperiodic arrangement of flux, respectively. Under the tight-binding formalism, the Hamiltonian reads as follows, 

\begin{equation}
	\begin{split}
	\mathcal{H} &= -t_{x}\sum_i \big(e^{i\theta_{n,n+1}}a_{n+1}^{\dagger}a_n+ b_{n+1}^{\dagger}b_n\big)- t_{y}\sum_n a_n^{\dagger}			b_{n}^{\dagger}\\
	&\quad  -\epsilon \sum_n (a_{n}^{\dagger}a_{n}+ b_{n}^{\dagger}b_{n}) + h.c.
	\end{split}
	\label{eq:ham}
\end{equation}

Here, $b_{n}^{\dagger} (b_n)$ [or equivalently, $a_{n}^{\dagger} (a_n)$ ] is the fermionic operator that creates (annihilates) a particle at the $n$-th site of the b-leg (or, a-leg). Hopping amplitudes $t_x$ and $t_y$ are the tunnelling matrix elements along the leg and along rung, respectively. When an electron hops in a closed loop from $r_n$ to $r_m$, the phase accumulated is

\begin{equation}
	\begin{split}
	\theta_{n,m} &= -\frac{e}{c\hslash} \int_{r_n}^{r_m}\vec{A}.\vec{dr} \\
	&\quad  = \frac{2\pi}{\Phi_o} \int_{r_n}^{r_m}\vec{A}.\vec{dr}
	\end{split}
\end{equation}

where $\vec{A}(\vec r)$ is the vector potential and $\phi_o$ is the flux quantum. The ladder is in the $xy$- plane and the magnetic field is applied in the $z$- direction. We choose the Landau gauge $\vec A= (-By, 0, 0)$ such that $\vec B= \vec{\triangledown} \times \vec A$.  Further, we exploit the gauge freedom and put the lower leg on the $y=0$ axis. As a result, the particle only adopts a phase when moving in the upper leg direction. Thus, the effect of the magnetic flux is realized through the aperiodically modulated Peierls' phase tagged to the hopping in the a-arm, viz, $t_x \exp (i\theta_{n,n+1})$ and its time-reversed partner. In our set up,  $\theta_n=2 \pi \Phi_n$ along the a-arm, and $\theta_n=0$ along the b-arm and the rung.
This choice of upper leg gauge does not alter the physical properties of the system~\cite{hugel,roux,zheng,qiao}. For the sake of simplicity, we set the onsite potential term $\epsilon$ to zero and consider the natural unit scale $\hslash=c=k_B=1$.

\section{The Spectrum and the Transport}
\subsection{The Hofstadter Butterflies}
In this section, we present the energy spectrum of a flux-modulated ladder network as a function of the modulation frequency $\alpha$ for different values of the flux strength $\lambda$. Similar to the AAH situation, we see that a variation in $\alpha$ produces  butterflies in the energy landscape. The butterflies arise from the interplay of two parameters $\alpha$ and $\lambda$. However, it should be noted that this modulation is in the distribution of flux in contrast to the original AAH model. To obtain the energy spectrum, we use exact diagonalization of the Hamiltonian for a system with $n=500$ atomic sites.

Turning on the magnetic field, we first set $\lambda=1/2$. With an overall uniform profile, the electronic spectrum looks continuous as shown in the Fig.~\ref{fig:butterfly}(a). One of the characterizing features of the energy bands is the existence of the band-gaps. Here, we see a small band-gap, symmetric around the centre. Moreover, the signatures of band-overlapping can be seen as well if one moves towards the band edge [Fig.~\ref{fig:butterfly}(b)]. Next, we set $\lambda=1$ and as a result, the central band-gap broadens and multiple mini band-gaps open up around the edges.
A prominent change in the spectrum is observed when $\lambda=2$ in a sense that each of the Landau levels breaks into smaller subbands and produces a self-similar fractal pattern in the energy spectrum. In the Fig.~\ref{fig:butterfly}(c), the fragmentation of the bands is evident as one moves away from the centre. This is similar to the `Hofstadter butterfly' and can be regarded as its variant. The self-similarity of the butterfly is discussed later. Interestingly, the result is similar to the AAH model where the butterfly appears once the intensity of the flux modulation potential is set equal to to $2$. Our findings show that, a quasi-1D system that can be approximated as a strip geometry of the Hofstadter model, is capable of producing a butterfly spectrum. However, the butterfly is destroyed the moment we turn on the `slowness' index ($\nu=0.8$, for example) as illustrated in  Fig.~\ref{fig:butterfly}(d).

\subsection{Density of states and the transmission characteristics}
\begin{figure*}[htpb]
	\centering
	(a) \includegraphics[width=0.6\columnwidth]{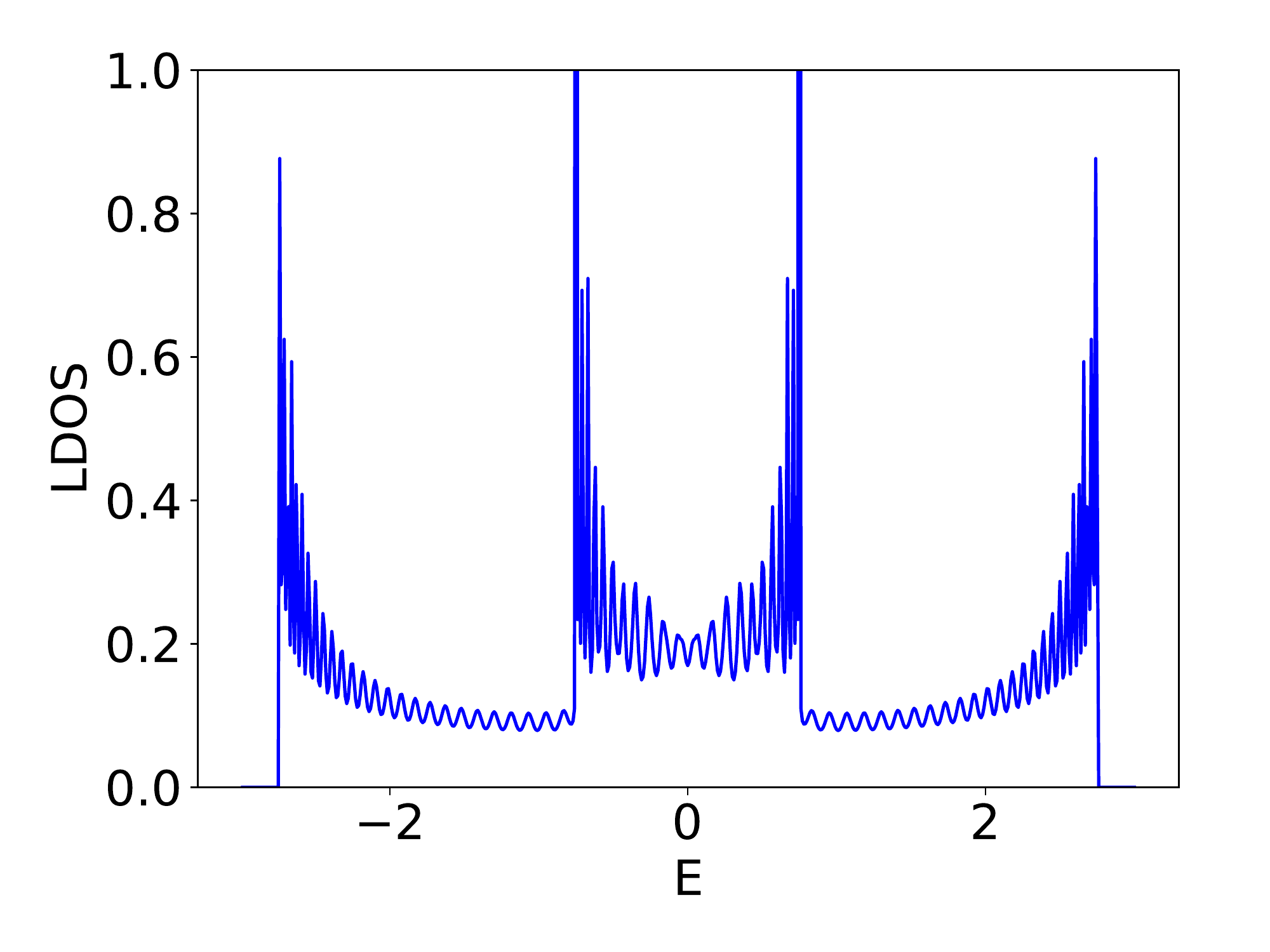}
	(b)\includegraphics[width=0.6\columnwidth]{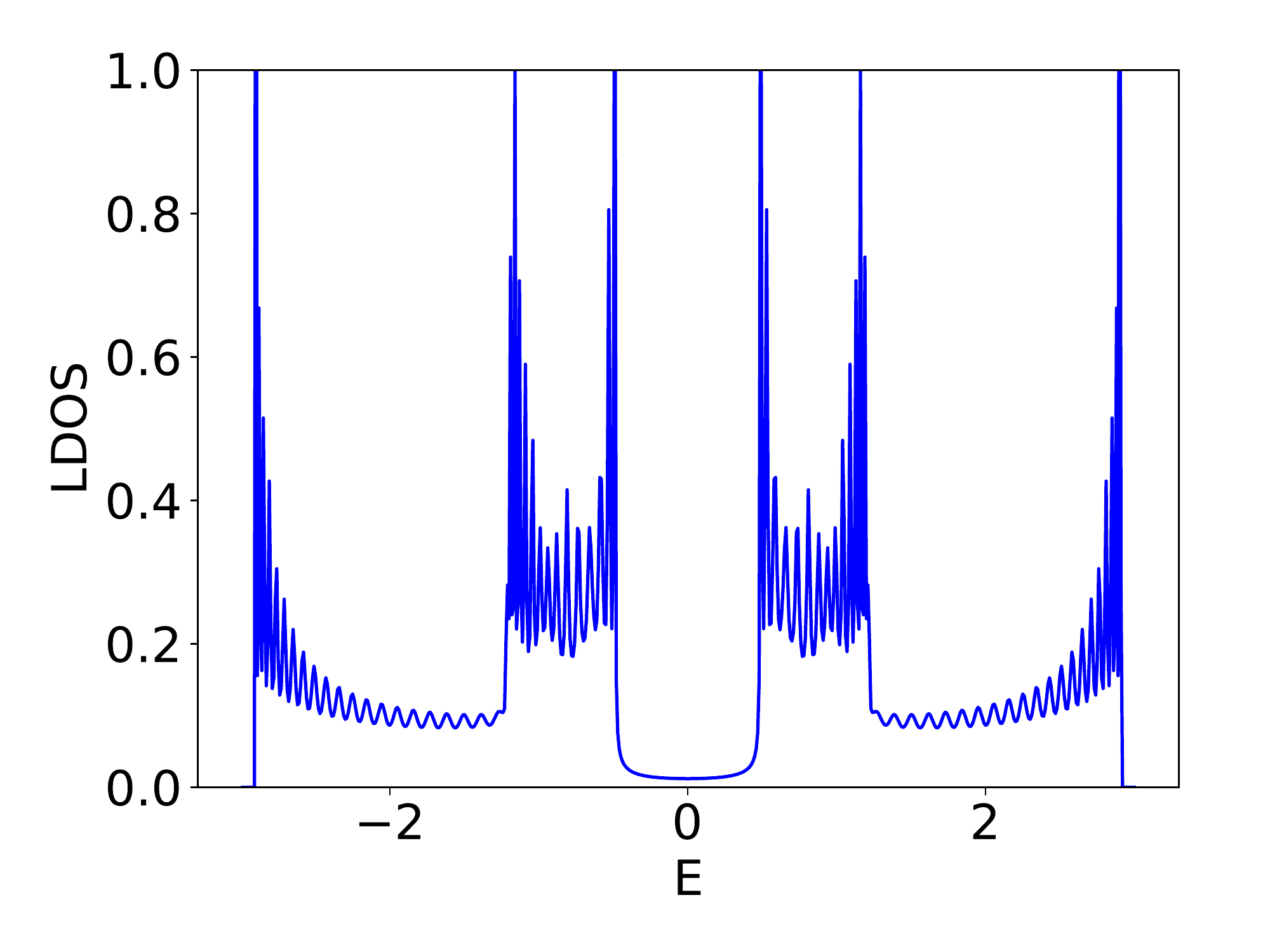}
	(c)\includegraphics[width=0.6\columnwidth]{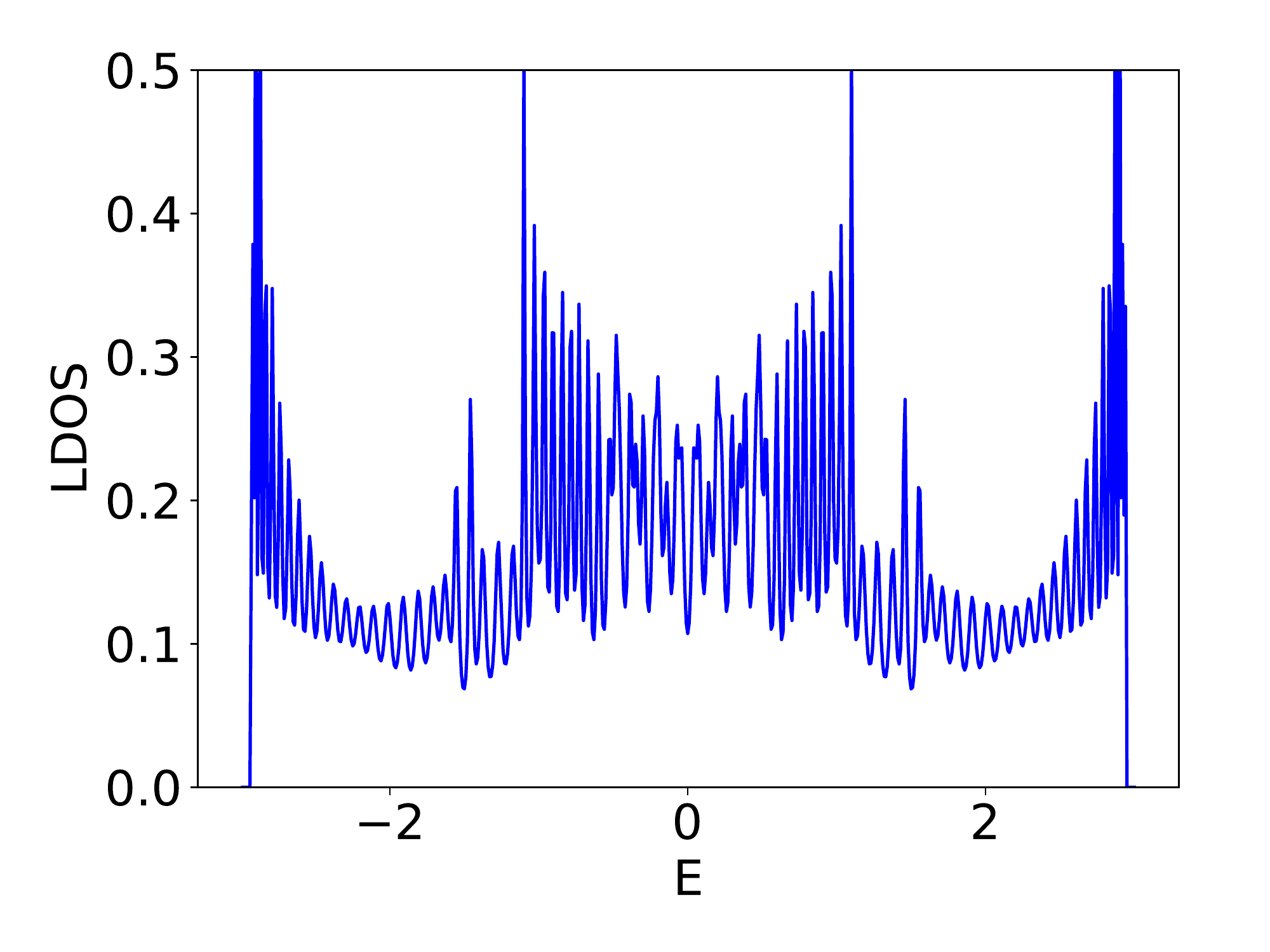}
	\\
	(d)\includegraphics[width=0.6\columnwidth]{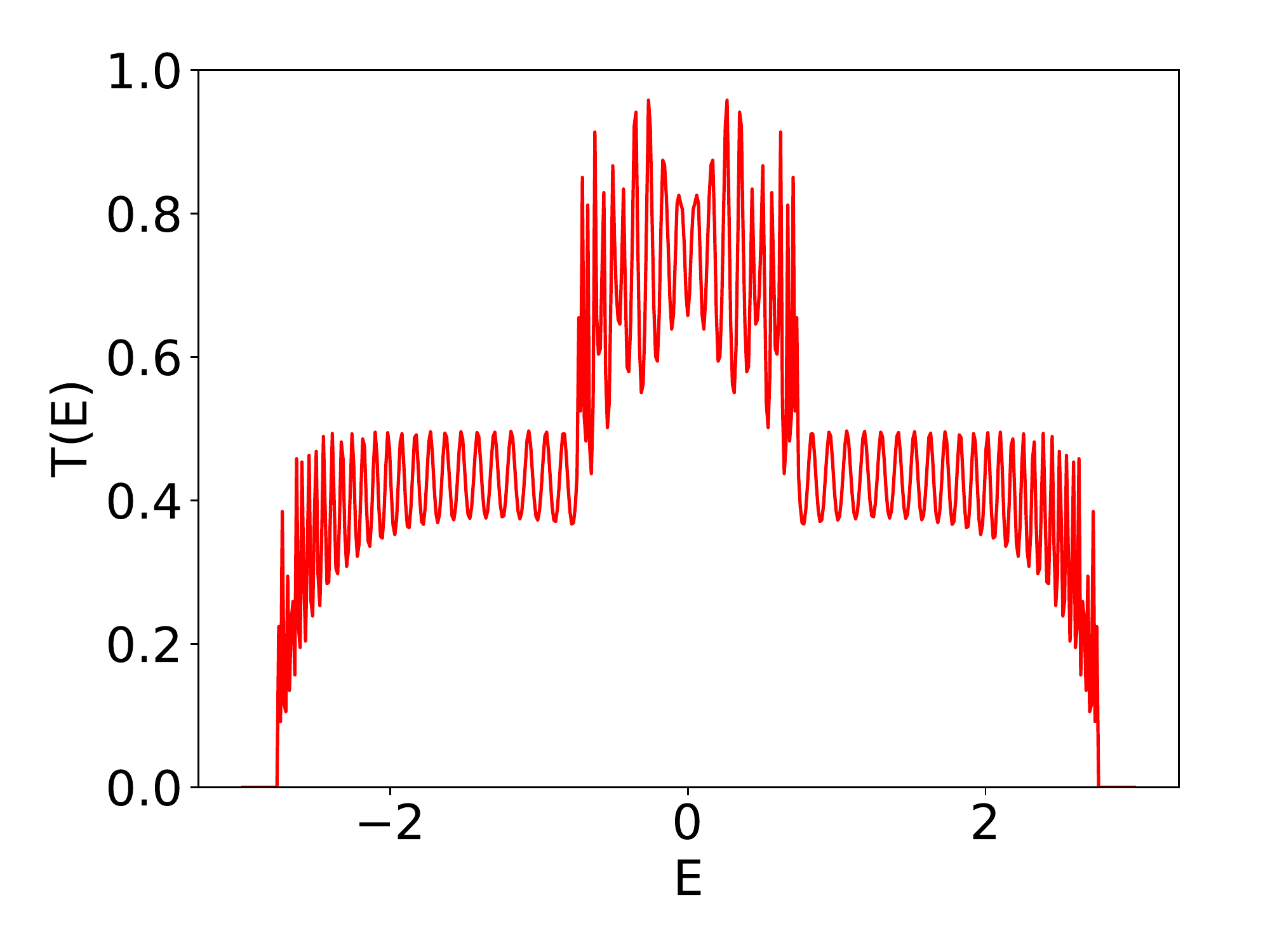}
	(e)\includegraphics[width=0.6\columnwidth]{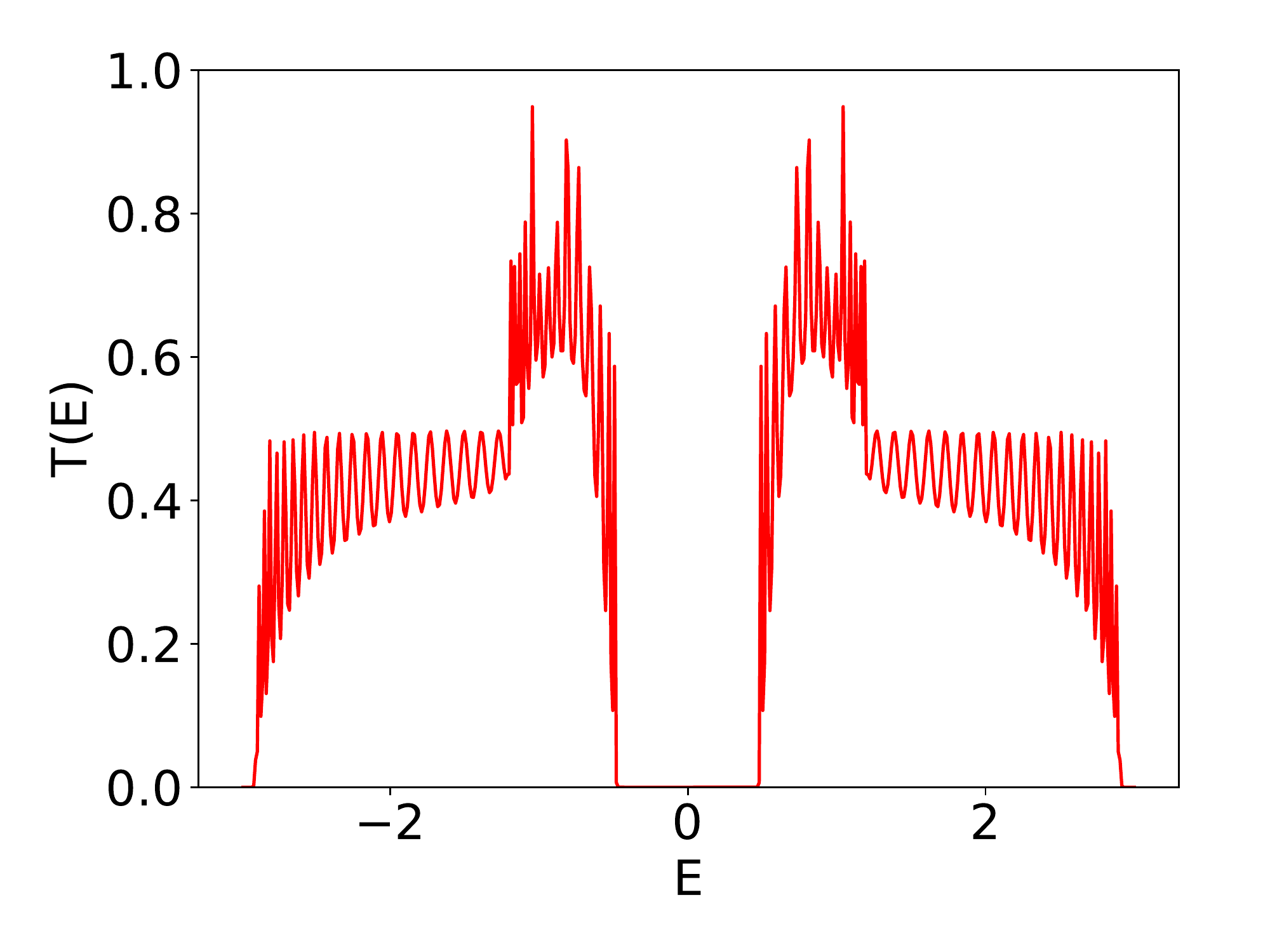}
	(f)\includegraphics[width=0.6\columnwidth]{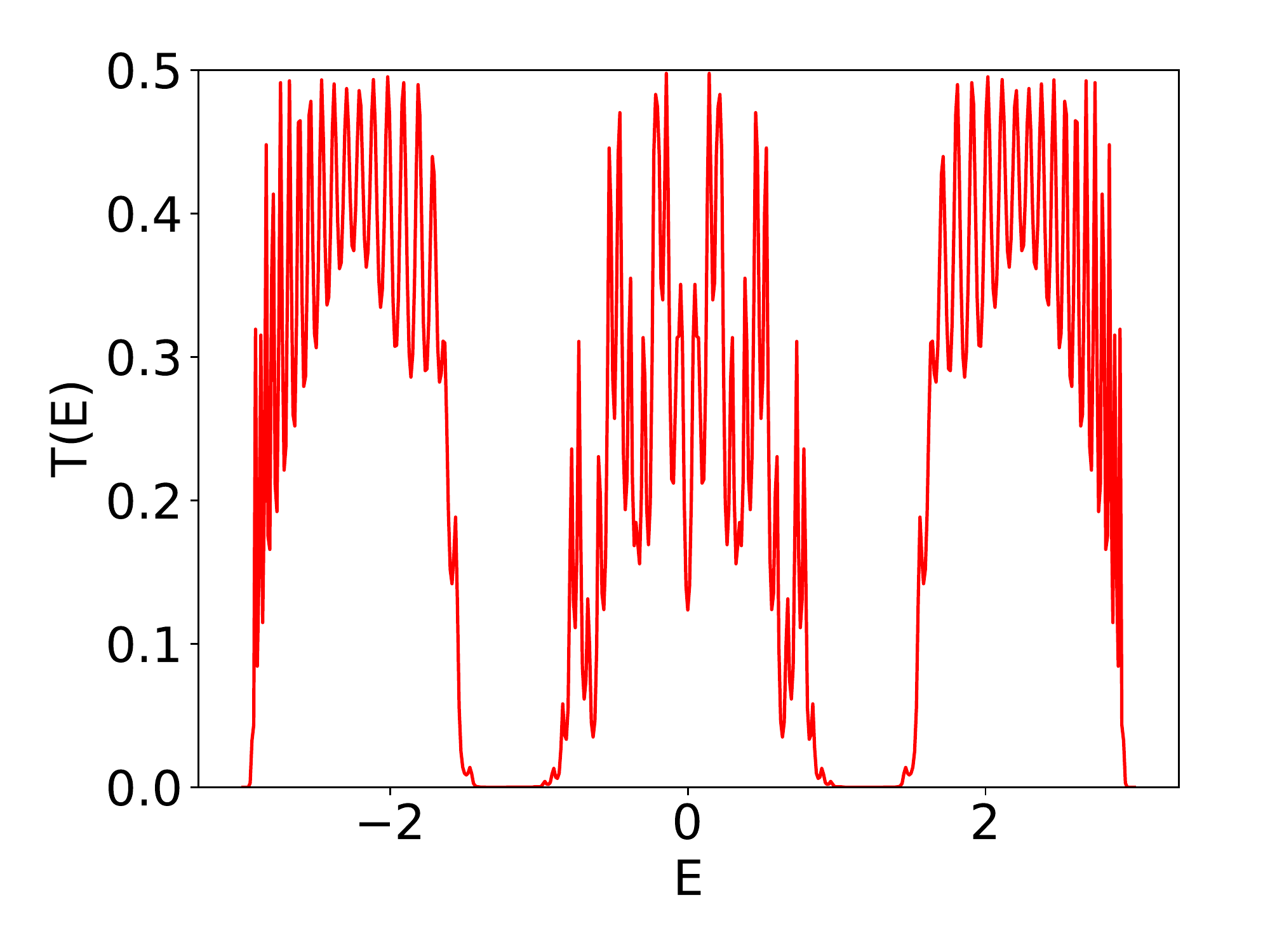}
	\caption{(Color online.) The upper panel shows the LDOS profiles of the two-strand ladder network with the leads. The transport characteristics are presented in the lower panels, just below the corresponding LDOS plots. We have set $\lambda=1$, and $t_x=t_y=1$. (a) and (d) correspond to a uniform flux distribution ($\alpha=1$, $\nu=1$). (b) and (e) correspond to a staggered flux distribution ($\alpha=1/2$, $\nu=1$) and a gap opens up in the spectrum. (c) and (f) depict the slow, aperiodically modulated case ($\alpha=\frac{\sqrt{5}-1}{2}; \nu=0.8$). Energy is measured in units of $t_x$.} 
	\label{fig:laddertransport}
\end{figure*}

To study the DOS and transport properties of a double-arm ladder, we adopt the two-terminal Green's function formalism. In this process, we attach two semi-infinite pair of conducting leads at the left and the right ends of the ladder~\cite{odashima} and present our results in the Fig.~\ref{fig:laddertransport}. Due to the coupling of the leads to the finite sized ladder, the Hamiltonian takes the form,

\begin{equation}
	\mathcal{H}= \mathbf{H}_{S}+ \sum_{l=1}^{4} \mathbf{H}_l+ \mathbf{H}_{lS}+ \mathbf{H}_{lS}^{\dagger}
\end{equation}

where $H_S$ is the tight-binding ladder (system) Hamiltonian and the term $H_{lS}(H_{lS}^{\dagger})$ accounts for the coupling between $l$th lead to the system $S$. The index $l$ runs over all the leads that are connected to the central part and can go up to $4$ in our case. With the help of Lodwdin’s technique~\cite{lowdin1,lowdin2}, we can easily partition the Hamiltonian and arrive at the expression for the effective Green's function of the system that reads

\begin{equation}
	\mathbf{G}_S= \big(E\mathbf{I}- \mathbf{H}_S- \sum_{l=1}^{4}\mathbf{\Sigma}_l \big)^{-1}
\end{equation}

where $\Sigma_l$ is the self-energy correction term that arises due to attachment of the lead to the system and contains all the information about coupling. The expression for the self-energy term of the lth lead is as follows

\begin{equation}
	\mathbf{\Sigma}_l= \mathbf{H}_{lS} \mathbf{G}_l \mathbf{H}_{lS}^{\dagger}
\end{equation}

Here, $\mathbf{G}_l= (E\mathbf{I}-\mathbf{H}_l)^{-1}$ corresponds to the Green's function matrix for the $l$-th lead. Once the self-energy is obtained, it is straightforward to calculate the coupling function, $\mathbf{\Gamma}_l (E)$, viz, 

\begin{equation}\label{coupling}
	\mathbf{\Gamma}_l (E)= i\big[\mathbf{\Sigma}_{l}^{ret}(E)- \mathbf{\Sigma}_{l}^{adv}(E)\big] 
\end{equation} 

In the above equation, $\Sigma_{l}^{ret(adv)}$ signifies the retarded (advanced) self-energy term. We make use of the fact that they are Hermitian conjugate to each other and rewrite the Eq.~\ref{coupling} as,

\begin{equation}
	\mathbf{\Gamma}_l (E)= -2Im\big[\mathbf{\Sigma}_{l}^{ret}(E)\big]
\end{equation}

The final expression for the transport between $m$-th and $n$-th lead, as a function of the system-lead coupling can be written as~\cite{cook,paramita,supriyo},

\begin{equation} 
	T_{mn}= Tr\big[\mathbf{\Gamma}_m \mathbf{G}_{S}^{ret}\mathbf{\Gamma}_n \mathbf{G}_{S}^{adv}\big]
\end{equation}

Here, $\mathbf{\Gamma}_{m(n)}$ denotes the matrix coupling the system to the $m$($n$)-th lead and $\mathbf{G}_{S}^{ret(adv)}$ is the matrix of the retarded (advanced) Green's function of the system. To minimize the error, we estimate transmittance for all four possible combinations of lead attachments and take an average of it. Therefore, the transport of a two-arm ladder is, 

\begin{equation}\label{trn_eq}
	T(E)= \frac{(T_{L_1R_1}+T_{L_1R_2})+(T_{L_2R_1}+T_{L_2R_2})}{2}
\end{equation}

\subsubsection{The local density of states}
The spectral canvas of a flux modulated two-arm ladder can be well understood through a  density of states profile, followed by a multi-terminal transport across a finite sized system. As mentioned earlier, the `twist parameter' $\alpha$ determines the flux modulation frequency as well as serving as a gateway to achieve different flux distribution in the plaquette. We try to understand how different arrangement of fluxes that decides the hierarchy of the twist angle affects the spectral scenario. 

The local density of states (LDOS) is obtained by calculating the Green's function of the lead-connected system, that is given as follows:

\begin{equation}
	\rho(E)= \lim_{\eta\to 0} \bigg[-\frac{1}{\pi N}Im\big(G_{00}(E+ i\eta)\big) \bigg]
\end{equation}
where $G_{00}(E+i\eta)$ is the diagonal element of the Green's function defined through the relation 
$\mathbf{G}_s = \big(E\mathbf{I}- \mathbf{H}_S- \sum_{l=1}^{4}\mathbf{\Sigma}_l \big)^{-1}$. The small pre-assigned quantity $\eta$ (set equal to $10^{-4}$ in the present calculation)  is added to the energy to avoid any singularities. The expression is normalized by the total number of atomic sites $N$.  $\mathbf{I}$ is the identity matrix having the same dimension as the system Hamiltonian matrix $\mathbf{H}_S$. The self-energy correction term $\Sigma_l$ appears due to the system-lead coupling. We have kept our energy scanning interval small enough to ensure a detailed distribution of the eigenstates.

We first study the spectral character with a uniform flux distribution. This is easily achieved by setting $\alpha=1$. The LDOS is presented in the Fig.~\ref{fig:laddertransport}(a),  where we restrict the presentation of the LDOS  within the value of $2$ just to bring out the smaller peaks. The ladder network used for the computation here has $n=100$ atoms. We have chosen $\lambda=\nu=1$. Such choices of the parameters ensure $\Phi_i/\Phi_0=1/2\pi$, a constant, yet a nominal flux piercing each plaquette. The time reversal symmetry along the upper arm is still broken, and the hopping there takes up the values $t_x \exp (\pm i)$ along the forward and the backward direction respectively on each bond in the upper arm. The uniformity of the phase is seen to result in a gapless spectrum, centered around $E=0$. The eigenstates in each case reported here bear an extended  character (we can rule out localization here as there is no intrinsic disorder anywhere in the system). The finite size of the network however causes an oscillating LDOS profile.  There is an interesting segmentation in the behavior of the two terminal transport, for which a plausible argument is provided in the subsequent discussion, and also in the Appendix A.

A major change in the spectrum is observed for $\alpha=1/2$ which corresponds to a staggered flux distribution. In the Fig.~\ref{fig:laddertransport}(b), the dense cluster of states around $E=0$  vanishes. Instead, we see the opening of a gap in the middle of the spectrum and a two subband pattern emerges in the LDOS profile. 

To study the effects of an aperiodic flux modulation we consider $\alpha$ to be irrational. Our choice here is $\alpha=\frac{\sqrt{5}-1}{2}$. Motivated by the works of Ref.~\cite{griniasty, dassarma}, we choose to take a non-zero value of the slowness exponent and set $\nu=0.8$. The LDOS profile shows a clustering of states around the centre and at the edges. However, there is no gap in the spectrum, unlike the staggered case. The nature of such states will be commented upon in the subsequent sections.

\subsubsection{The transport characteristics}

We have computed the transmission characteristics for various flux distribution using  Eq.~\ref{trn_eq} and present the numerical results in the lower panel of Fig.~\ref{fig:laddertransport} to benchmark it with the LDOS results. The result for a uniform flux arrangement shows a resonating feature for an incoming electron with $-1\leq E\leq 1$ that leads to high transmittance. Beyond the above energy range and on either side of it, the transmission is lower, making a step-like appearance [see Fig.~\ref{fig:laddertransport}(d)] that is well known to exist for a ribbon like network, where an $M$ strand network gives rise to a total of $2M$ steps in $T(E)$~\cite{stegmann} . While the exact reason for a low or a high transmission coefficient at any particular energy is hard to justify in a quantitative manner (apart from making a qualitative statement about partially destructive or constructive quantum interference in the looped structure of the ladder network), a plausible explanation for the segmentation of the band can be obtained by looking at the difference equations in a rotated basis,  at least in the special case of zero flux. Such an argument is provided in the Appendix A.

Switching to a staggered flux, we immediately notice the extinction of the central conducting regime as shown in the Fig.~\ref{fig:laddertransport}(b) and (e). This zero transport region can be understood from the LDOS profile which shows that there are no energy states to offer any transmission. Beyond this central energy regime, there exists a low transport region. 

Finally, we focus on the SVF case where the flux arrangement is incommensurate with the underlying lattice periodicity. The average transport is low, compared to the earlier cases. In Fig.~\ref{fig:laddertransport}(c) and (f), we present the results and immediately observe that there are energy regimes where the LDOS (broadened by the attached leads) is non-zero, and yet the transmission coefficient vanishes. 
This implies that the states have become either critical or localized with a localization length falling well short of the system size. Its thus tempting to conjecture a metal-insulator transition. However,  to gain a better insight and to resolve the issue we resort to an analysis of the inverse participation ratio (IPR) and subsequently, to a study of the multifractality of the spectrum.

\section{Statistical analysis: The inverse participation ratio (IPR) and the multifractality}
\subsection{ The IPR and the nature of the eigenstates} 
\begin{figure}[h]
	\centering
	\includegraphics[width=0.6\columnwidth]{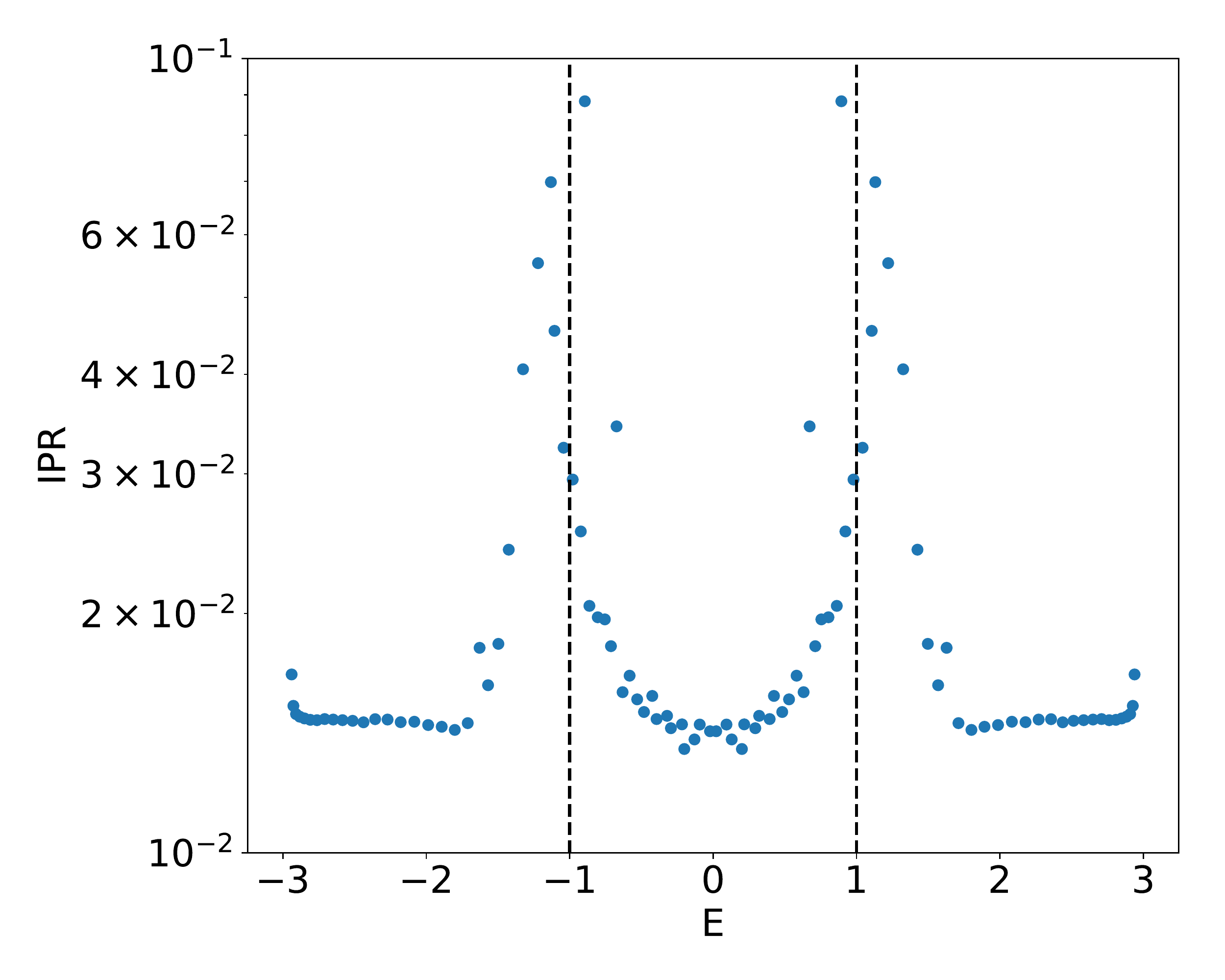}
	\caption{(Color online.) Inverse participation ratio of the whole spectrum is plotted against energy for the system with N=100 atomic sites. The parameters are $\lambda=1, \alpha=\frac{\sqrt{5}-1}{2}, \nu=0.8$.}
	\label{fig:ipr}
\end{figure}

We have estimated the inverse participation ratio (IPR) for the slowly varying quasiperiodic arrangement of flux. IPR is a good numerical tool to characterize localization. It is defined as the fourth power of the normalized wavefunction and is given by~\cite{tong,izrailev,flores},

\begin{equation}
	IPR \sim \sum_n |\psi_n|^4
\end{equation}

,where n goes over all the atomic sites. The IPR estimates how many numbers of sites are occupied by the wavefunction, and is defined as the inverse of this number. One of the reasons to use IPR is that it does not assume the exponential localization of the wavefunction beforehand and is very much sensitive to the nature of the wavefunction making it a good candidate to study the localization properties of a system. In  Fig.~\ref{fig:ipr}, we have plotted the IPR of the whole spectrum against energy. Interestingly, the IPR shows a dramatic increase around $E=\pm 1$, where the transport is pretty low.

We further investigate the minute details and find that,  this abrupt change in the value of IPR becomes sharper as the system size $N\rightarrow \infty$.  On the other hand, there are almost flat and  low IPR regions, close to the edges primarily, and at the center,  that represent the {\it delocalized} eigenstates, leading to a larger (sometimes resonant) transport profile. We notice the high IPR states differ almost by an order of magnitude compared to the low IPR counterparts. This asymmetry in the IPR suggests the co-existence of different quantum phases in the spectrum. A more rigorous approach is taken in later section to clarify the nature of the states. Nevertheless, it is safe to say this result at least presents the possibility of a single-particle mobility edge - a domain wall specified by a particular energy that separates the localized (insulating) states from the extended (conducting) ones. It should be noted that the original AAH model shows a metal-insulator transition only in the parameter space, and does not have any mobility edge since all the states are either delocalized or localized depending upon the value of the strength of the on-site potential~\cite{sokoloff}

\subsection{Multifractality of the spectrum}
Multifractal behaviour is a signature feature of strongly fluctuating wavefunctions at criticality. Unlike monofractals, multifractal systems are defined by continuous set of exponents illustrating the scaling of moments of some probability distribution. For our case, this probability measure is $\big| \psi(r) \big|^2$. Now we can define generalized IPR ($I_q$) as the moments of eigenstate intensities~\cite{evers,hiramoto2},

\begin{equation*}
	I_q= \int  \big| \psi(r) \big|^{2q} d^{d} r
\end{equation*}

At criticality, $I_q$ follows a power-law relation with the system size $N$, $I_q \sim N^{-\tau(q)}$. Here, $\tau(q)$ is a continuous set of exponents that characterizes a multifractal system.  The normalization of the wavefunction requires $I_1= 1$. With this constraint, one can write  $\tau (q)$ as $\tau (q)= (q-1)D_q$, where $D_q$ is the generalized fractal dimension. It should be noted that the critical exponent $\tau (q)$ $(i)$ is linear with $D_q=d$ for metals, $(ii)$ becomes flat in localized systems with  $D_q=0$ and, $(iii)$ is a non-linear function of $q$ at critical points. 

\begin{figure}[h!]
	a)\includegraphics[width=0.6\columnwidth]{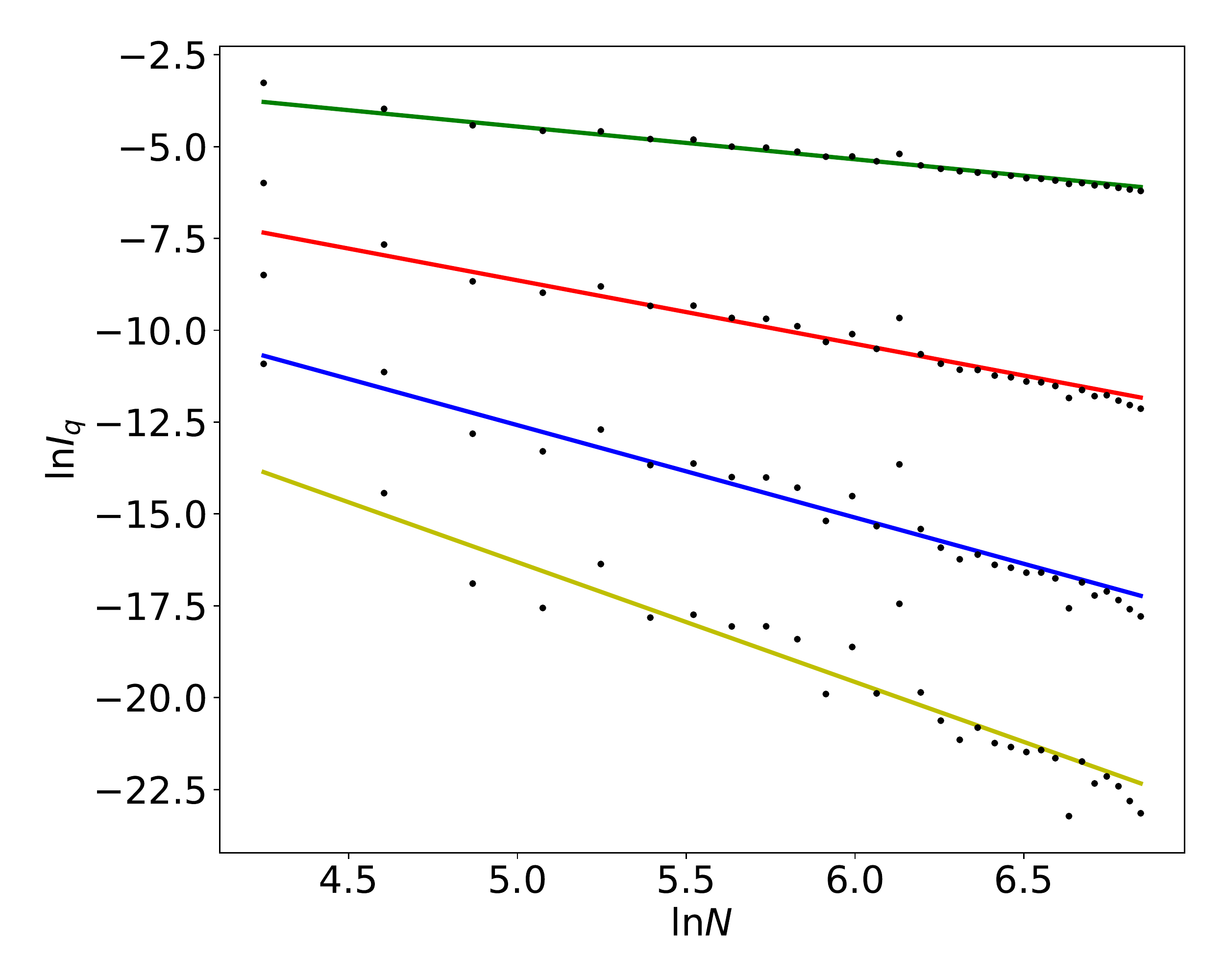}
	b)\includegraphics[width=0.6\columnwidth]{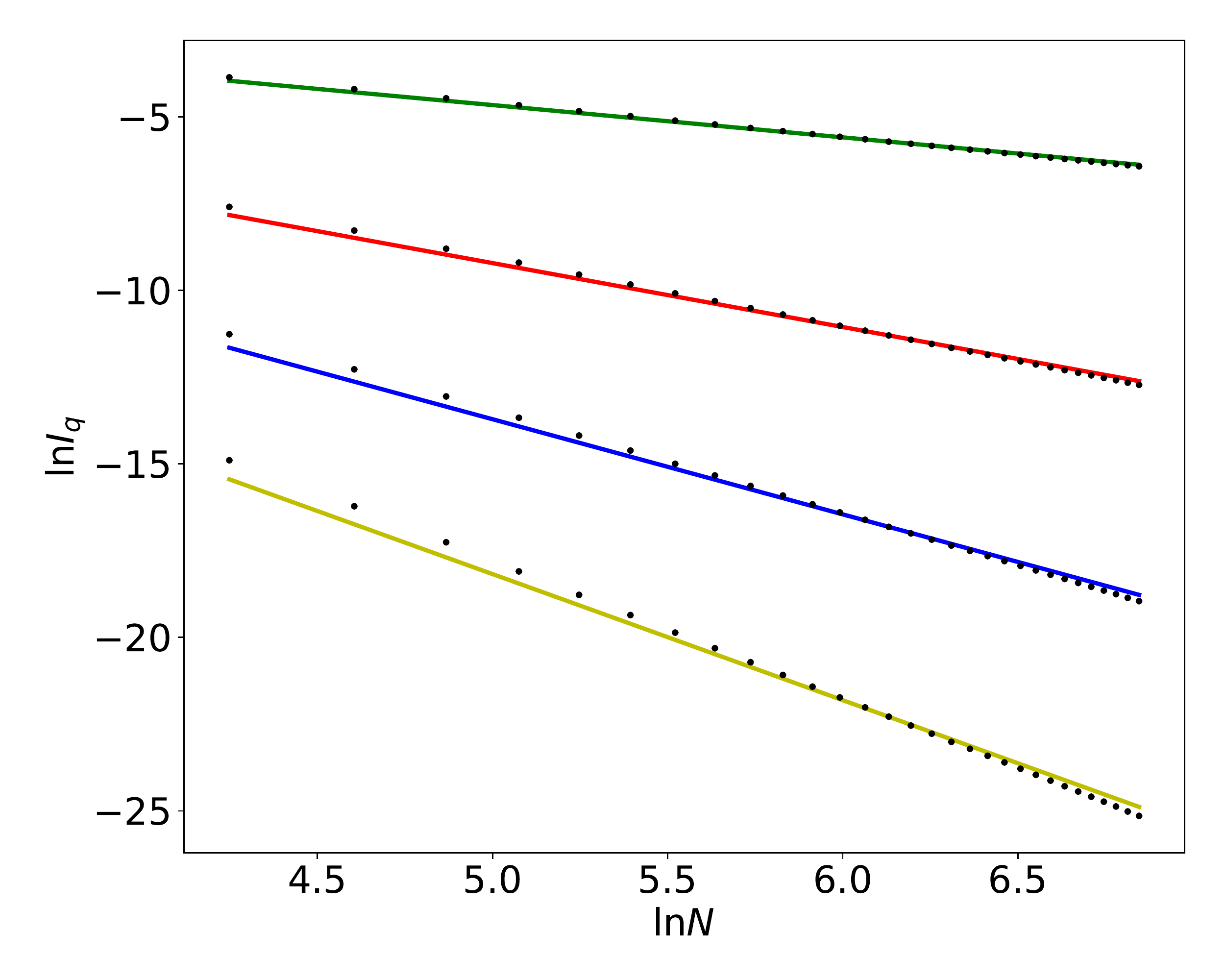}
	c)\includegraphics[width=0.65\columnwidth]{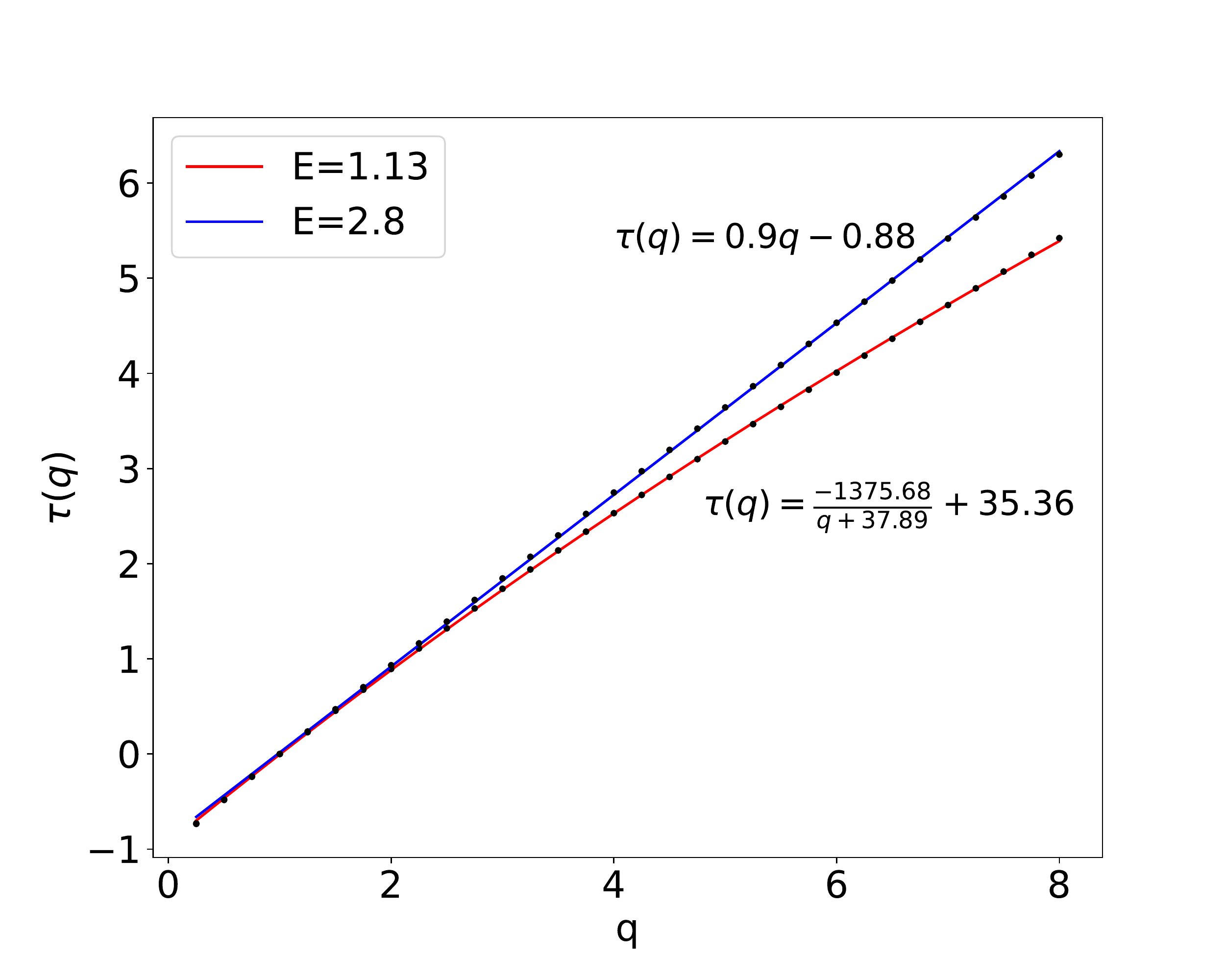}
	\caption{(Color online.) Plot of $\ln I_q$ versus $\ln N$ is shown for two energies  a) $E=1.13$,   b) $E=2.8$, and c) with system size $N$ ranging 	from 40 to 1050. Here, q is a parameter and takes the value 2(green), 3(red), 4(blue), and 5(yellow). c) The trend of multifractal exponent $\tau (q)\sim -\frac{\ln I_q}{ln N}$ against $q$ demonstrates the 		difference between extended and critical phase.}
	\label{fig:mft}
\end{figure}

As seen earlier, our results exhibit two different regimes of phases in a single spectrum for a slowly varying aperiodic sequence of flux. Here, we unfold the nature of the underlying quantum phases with the standard multifractal analysis. This is a technique that has been successfully applied in many disordered and quasiperiodic models~\cite{vasquez,godreche,hiramoto,liu}. To do this we set $\alpha= \frac{\sqrt{5}-1}{2}; \nu=0.8$ and pick two suitable eigenenergies  $E=1.13$, and $2.8$ (in units of the hopping integral $t_x$) from the high and low regions of IPR respectively. We also make sure that both of the energies are present at any system size. The plot of the generalized IPR versus system size highlights the difference between the two phases [see Figs.~\ref{fig:mft}(a), (b)]. For the {\it critical} states ($E=1.13$), $I_q$ changes non-linearly against $N$ with the effect becoming sharper at higher values of $q$ while, on the contrary, the trend is linear at all $q$ for metallic phase($E=2.8$). We calculate the multifractal exponent by plotting $I_q$ versus $N$ at different $q$ ranging from 0 to 8 in an interval of 0.25 [see Fig.~\ref{fig:mft}(c)]. The blue linear fit demonstrates the metallicity for $E=2.8$. At $E=1.13$, $\tau (q)$ changes non-trivially with q and deviates more and more as the $q$ increases- an inherent feature of the multifractality. A non-linear fit shows $\tau (q)$ varies as $\approx \frac{-1375.68}{q+37.89}+35.36$ which explains the effect of non-linearity becoming more prominent at higher $q$.

\section{Topological edge states}
With the discovery of topological insulators (TI)~\cite{tknn,wen,bernevig}, the concept of topology have become essential in understanding electronic transport. A striking feature of topological insulators is that they host novel edge states on their surface. Such states are conducting in nature and move around the insulating bulk without being affected by any disorder present in the system. The existence of an edge state is defined by the open boundary of the system, and its dynamics is independent of any microscopic details. This fact makes such systems an excellent choice for fabricating disorder-free transport technology. The path breaking Su-Schreiffer-Heeger model for example, provides an excellent platform to understand the occurrence of the edge states and a topological phase transition~\cite{ssh}.
\begin{figure}[h!]
	\centering
	(a)\includegraphics[height=7cm,width=7cm]{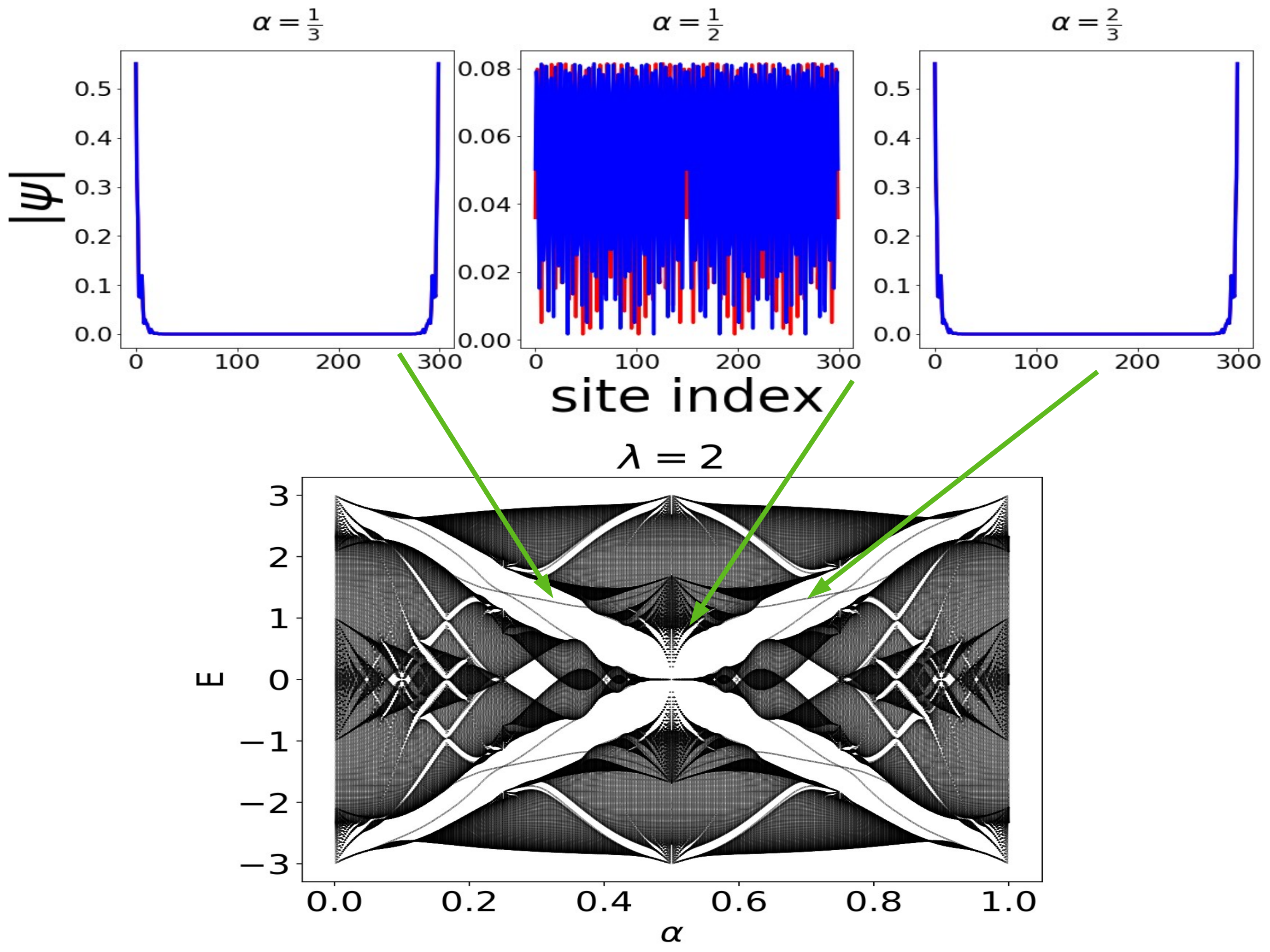}
	(b)\includegraphics[width=0.7\columnwidth]{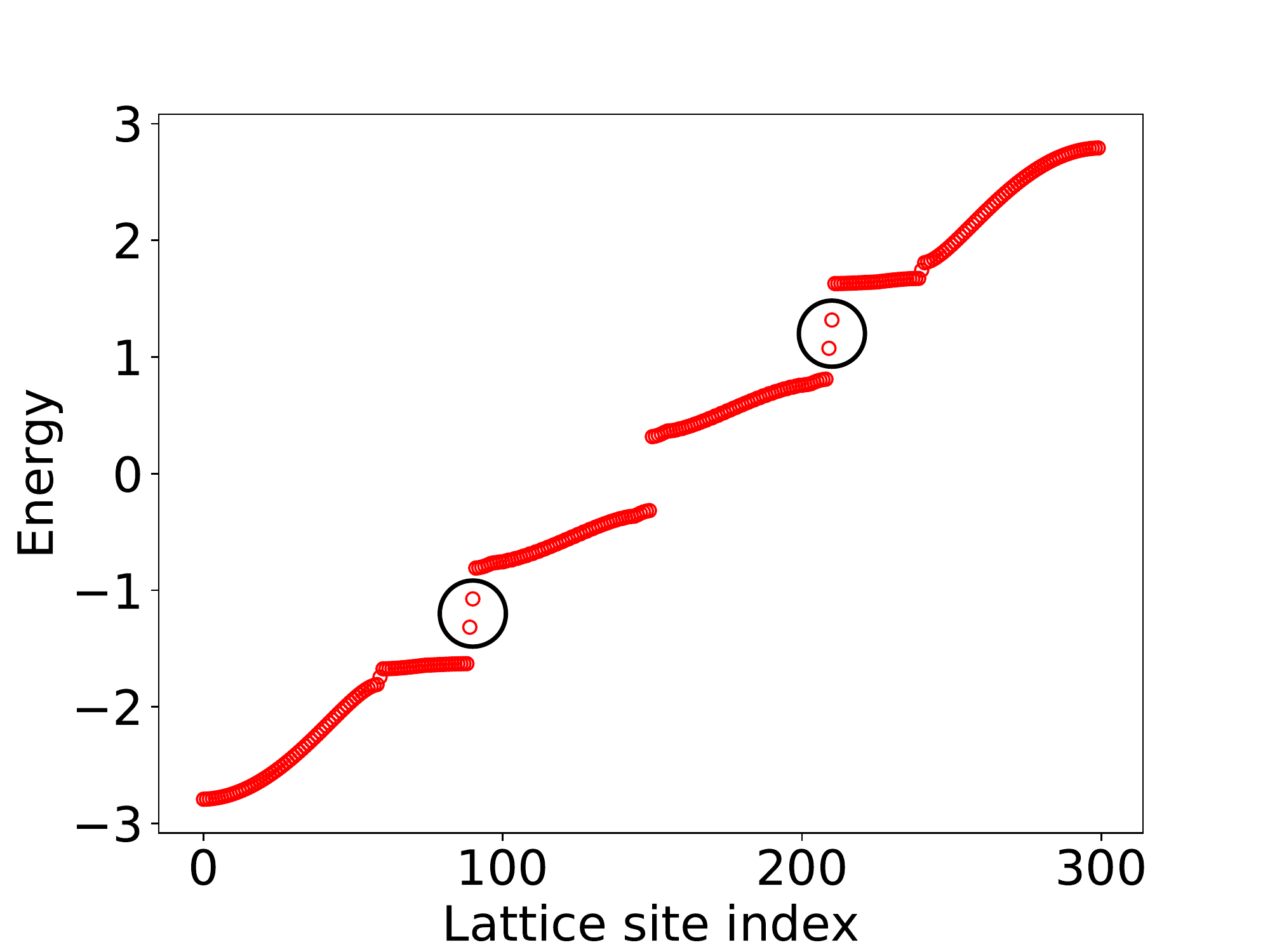}
	(c)\includegraphics[width=0.7\columnwidth]{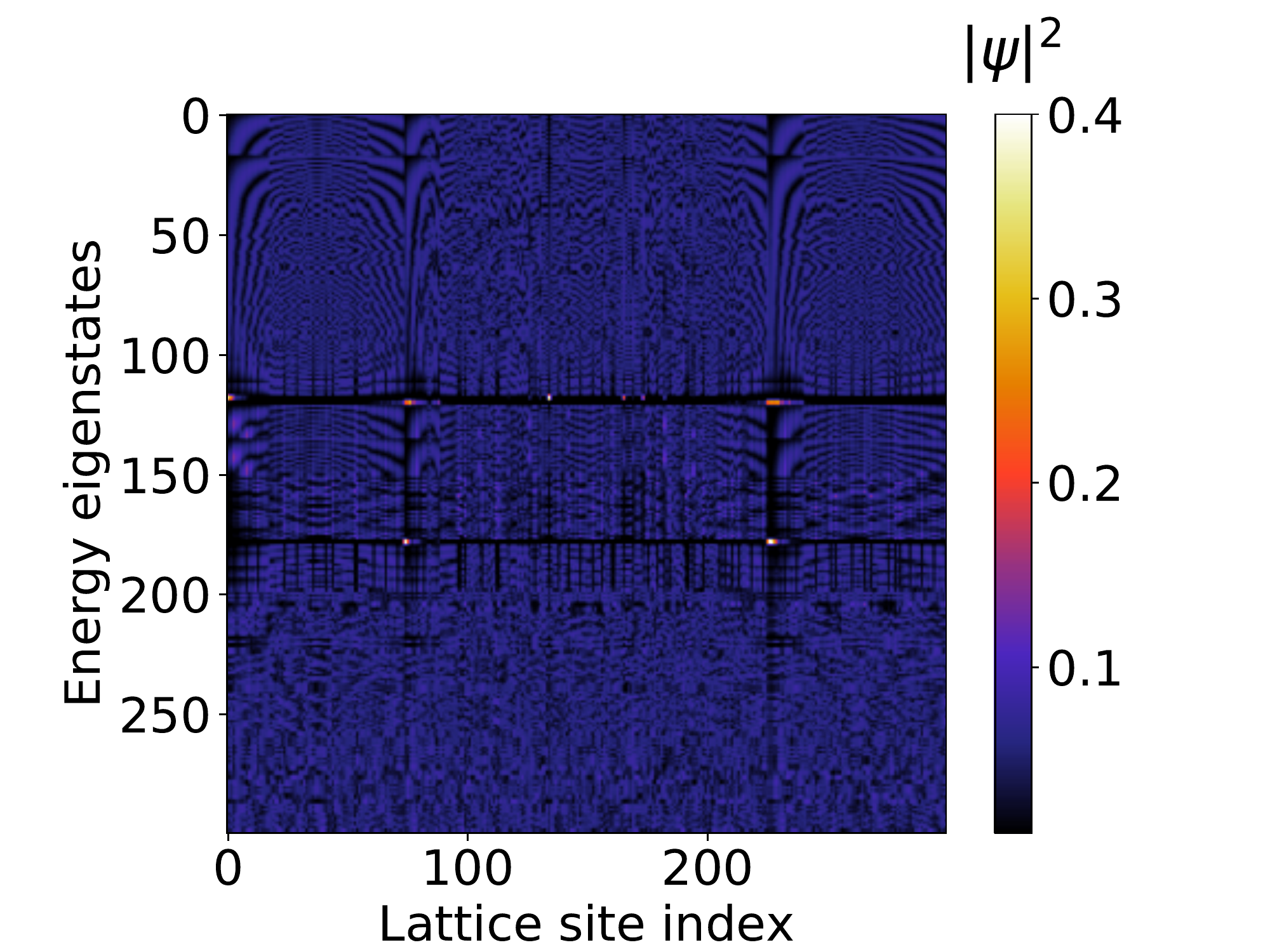}
	\caption{(Color online.) a) A butterfly energy spectrum under open boundary condition is shown to demonstrate the traces of the topological edge states. The Hamiltonian parameters are $\lambda=2, \nu=1,t_{x,y}=1$. Both the gapped states are found to be localized at the edges except for the bulk region as shown by the amplitude versus position plot above. Next we analyze an eigenstrip of the Butterfly spectrum that contains the gapped states, and fix $\alpha=\frac{1}{3}$. In (b) a plot of the energy versus lattice site index is shown. Traces of the edge states can be infered from the black circles that highlight the bulk-isolated energy points, and in  (c) the probability density $|\psi|^2$ of the entire eigenspectrum against lattice site index unfolds the spatial and intensity distribution of the states across the ladder. Four bright spots indicate that the two pairs of edge modes must be localized in nature.}
	\label{fig:edge}
\end{figure}

To elucidate the nontrivial topological states, we study a generic butterfly spectrum under open boundary condition. Two pairs of edge states can be seen in the gapped region of the spectrum in the Fig.~\ref{fig:edge}(a). We pick up three states from different sections of the spectrum, for $\alpha= 1/3,1/2$ and $2/3$,  and plot the amplitude of the wavefunction against the position in each case to highlight the contrast between the bulk and the gapped states. Out of the three, the two gapped states at $\alpha=1/3$,and $\alpha=2/3$ are predominantly stationed at the interface, while for $\alpha=1/2$,  in the bulk, the energy is distributed all over the Bloch bands with a non-zero amplitude everywhere. 

In Fig.~\ref{fig:edge}(b) we see the same picture in the energy versus site index plot with two pairs of edge states isolated from the rest of the spectrum. The right panel shows the density plot of the probability distribution for the whole spectrum [Fig.~\ref{fig:edge} (c)]. Energy eigenstates are plotted in the $Y$-direction against the lattice site index on the $X$-axis while the color bar showcases the probability density of wavefunction $\big| \psi \big|^2$. We mostly see a shade of dark-violet pattern prevailing everywhere except at the four bright points. These high-intensity points indicate that the edge states are localized in nature and must live in the vicinity of the physical edge of the ladder.

\begin{figure}[h!]
	\includegraphics[width=0.7\columnwidth]{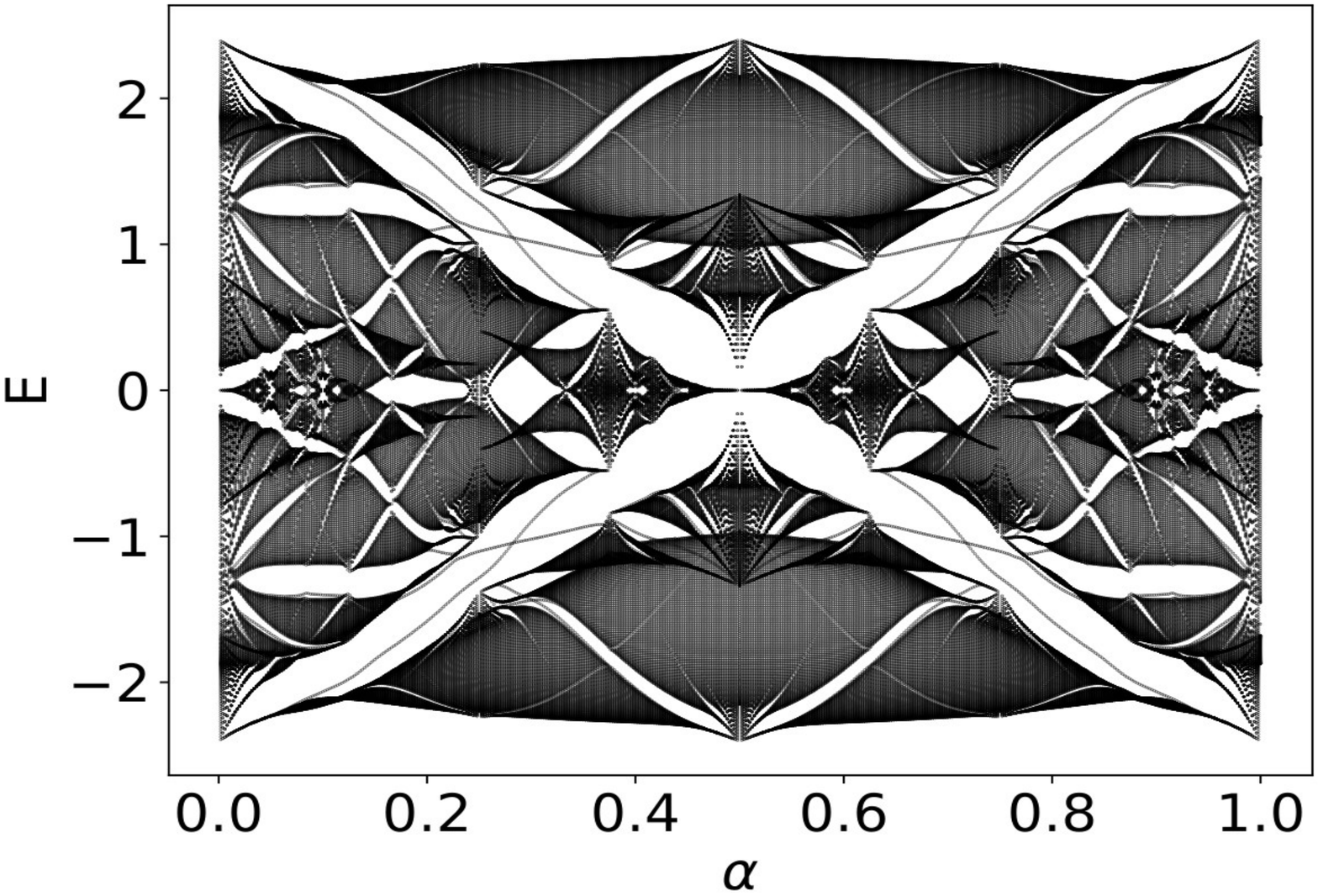}
	\caption{The electronic spectrum of a generic butterfly under hopping perturbation is shown. Here, the hopping elements are $t_{x,y}^{n+1}=1,t_{x,y}^{n}=0.8$. All the edge modes survive the perturbation suggesting their robustness against local distortion. }
	\label{fig:robustness}
\end{figure}

One often associates the term `robustness' in the context of edge states to describe the extent to which a state survives under a local distortion~--- a striking feature of the novel insulators arising from the non-trivial topology of the energy bands. There of course exists a wide variety of techniques in the current literature each serving a different need of individual systems to check the robustness. In this work, we work with a staggered distortion in the hopping element to test the fate of the edge states under such disturbance. The hopping element definitely depends, apart from other issues, on the distance between two adjacent atomic sites. Thus, by employing a perturbation we can mimic the physical equivalence of stretching or squeezing a ladder. For this purpose, we introduce a distortion amplitude $\delta \in [0,1]$ so that the hopping parameters take the form $t_{x,y}^{n+1}=t, t_{x,y}^{n}=t-\delta$ in the $(n+1)$th and the $n$-th plaquette respectively. With $t$ set equal to $1$, we observe that increasing the amplitude of fluctuation $\delta$ does not affect the edge states in any appreciable manner as long as $\delta$ is not too high (e.g. beyond $0.5$). We present our result of the butterfly spectrum, shown in Fig.~\ref{fig:butterfly}c, under perturbation for a median value of $\delta=0.2$ [see Fig.~\ref{fig:robustness}]. As expected, the hopping asymmetry opens up multiple band-gaps in the spectrum. The most interesting part is that all the pairs of the edge states survive the presence of distortion for amplitude as high as $\delta=0.2$. This demonstrates that, any local change such as the physical distortion of the ladder does not affect the edge states. 

The topological origin of these states can be traced back to a chiral symmetry of the ladder Hamiltonian. This has been discussed in details in Appendix B, considering the specific case of $\alpha=1/3$, and Fig.~\ref{fig:edge}. The argument can easily be extended to other values of $\alpha$ to ascertain any chiral symmetry connection. Interestingly, very recently, higher-order topological insulators described by AAH potentials have been reported to exhibit a similar chiral symmetry protection of the edge states~\cite{zeng}. 

\section{Possible experimental setup}
We propose that an optical lattice set-up using ultracold atoms may  realize our work experimentally. Ultracold atoms in optical trap provide an excellent opportunity to test different theoretical models in many areas of physics as diverse as quantum information~\cite{yang}, statistical physics~\cite{renzoni} and high-energy physics~\cite{schoop}. The process involves interfering several counter-propagating laser beams to artificially create a controlled and tunable periodic light system where atoms can be trapped. As shown in the Ref.~\cite{li}, a $2D$ double-well potential is considered for fabricating a two-arm ladder network. High-controllability and precision of this method allow us to tune the well-depth by varying the ratio of the laser amplitudes and the phase between them. However, our system requires the use of charged particles. Thus we suggest incorporating synthetic gauge fields to overcome the charge-neutrality of the atoms. Synthetic gauge fields are achieved by a pair of Raman lasers that can imprint a complex phase, analogous to Peierls' phase on the wavefunction as it hops between atomic sites~\cite{lin,dalibard,celi}. The ability to tune the lattice constant paired with the artificial gauge is useful to achieve magnetic flux as high as the flux quantum. This allows us to test theories without the need of astronomically large magnetic field which was not possible in conventional condensed matter lattices. Experimentally, strong  staggered~\cite{aidelsburger2011,struck} and uniform~\cite{aidelsburger2013} magnetic flux arrangements have been achieved in this framework and very recently, Macini et al.~\cite{macini} have reported to measure chiral edge current of a ribbon Hofstadter geometry within a synthetic gauge field setup.

\section{Conclusion} 
In summary, we have considered a two-arm tight-binding ladder network, defined by Eq.~\eqref{eq:ham}, in the presence of an external magnetic field. The key feature of our work is the inclusion of a cosine modulated magnetic flux that mimics a virtual axial twist of the ribbon structure. Such spatial variation of flux paired with low-dimension ladder network provides an excellent framework to study systems from ribbon Hofstadter geometry to twisted DNA network. We show that ladder networks are capable of producing butterfly pattern for a magnetic flux of few flux quantum. However, if the variation of flux is too slow the interference pattern dies out. Our main result indicates the co-existence of extended-critical phase for a slowly varying flux incommensurate with the underlying lattice. Furthermore, we use the multifractal analysis to investigate and characterize the nature of the phases. A significant change in the IPR along with the multifractal exponent calculation supports our claim of different phases. This result is of substantial importance, because the co-existence of quantum phases in the spectrum hints towards the existence of mobility edge. For a commensurate sequence of flux, the ladder hosts two pairs of non-degenerate topological edge states localized near the physical edge of the ladder. We propose a scheme of correlated hopping perturbation to test the robustness of edge states. We notice all of the edge modes are robust against the hopping perturbation up to a mark indicating their non-trivial topological origin. Finally, we suggest the use of assisted Raman lasers in an optical lattice with ultracold atomic configuration to realize our work in an experimental framework.

\vskip 0.2in
\textbf{Acknowledgement}
SS thanks Amrita Mukherjee for her help with the calculation of Green's function and helpful suggestions at times. SS is supported through the INSPIRE fellowship from the Department of Science and Technology, Govt. of India.

\appendix
\section{On the two-terminal transport across a ladder network}
Let us consider the clean case, with zero flux, to understand the heuristic argument that we wish to put forward.

With $\Phi_n=0$ (that is obtained by setting $\lambda=0$), the difference equations satisfied by the amplitudes of the wave functions $\psi_{n,a}$ and $\psi_{n,b}$ are easily written down as,
\begin{eqnarray}
(E - \epsilon) \psi_{n,a} & = & t_x (\psi_{n-1,a}+\psi_{n+1,a}) + t_y \psi_{n,b} \nonumber \\
(E - \epsilon) \psi_{n,b} & = & t_x (\psi_{n-1,b}+\psi_{n+1,b}) + t_y \psi_{n,a}
\label{app-1}
\end{eqnarray}

A change of basis~\cite{chakrabarti,chakrabarti2}, decouples the set of Eq.~\eqref{app-1} into,

\begin{figure}[h!]
	\includegraphics[width=0.9\columnwidth]{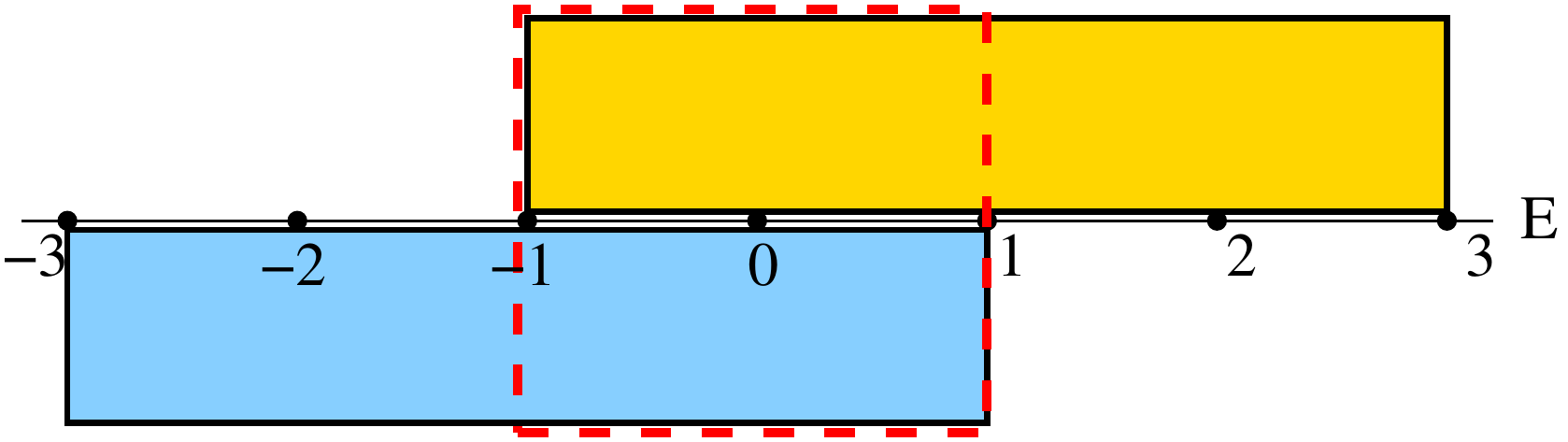}
	\caption{(Color online.) The schematic diagram that shows the extents of the energy bands of the decoupled ladder strands described in the new ${\phi_n}$ basis. The region of overlap is outlined by the red dashed lines.}
	\label{band}
\end{figure}

\begin{eqnarray}
\left [E - (\epsilon+t_y) \right ] \phi_{n,a} & = & t_x (\phi_{n-1,a} + \phi_{n+1,a}) \nonumber \\
\left [ E - (\epsilon-t_y) \right ] \phi_{n,b} & = & t_x (\phi_{n-1,b} + \phi_{n+1,b}) 
\label{app-2}
\end{eqnarray}
where, $\phi_{n,a} = (\psi_{n,a}+\psi_{n,b})/\sqrt{2}$, and $\phi_{n,b} = (\psi_{n,a}-\psi_{n,b})/\sqrt{2}$. The new, `rotated' $\phi$-basis represents an excitation that is well described by a linear combination of the old amplitudes. It should be noted that if any one of the amplitudes $\psi_{n,a}$ or $\psi_{n,b}$ is {\it extended} in nature, the combination naturally represents an extended wave propagation. On the other hand, in order that the $\phi_n$'s represent a localized state, both $\psi_{n,a}$, and $\psi_{n,b}$ have to be localized in space.

Each equation in the set of Eq.~\eqref{app-2} gives rise to absolutely continuous bands, ranging between $(\epsilon+t_y-2t_x$, $\epsilon+t_y+2 t_x)$, and 
$(\epsilon-t_y-2t_x$, $\epsilon-t_y+2 t_x)$. With $\epsilon=0$, and $t_x=t_y=1$, the bands coming from each equation are schematically shown in Fig.~\ref{band}. The region of overlap is put inside a red-dashed box. 

For a continuous distribution of energy between $E \in [1,3]$, and between $E \in [-3,-1]$ we come across the lower value of the transmission coefficient, that turns out to be approximately half the value $T(E) \simeq 1$ which spans the central energy regime $E \in [-1, 1]$. This is the result shown in Fig.~\ref{fig:laddertransport}(d), and not for $\Phi_n=0$ really. Our argument, given above, about the band-splitting, though assumes zero flux penetration, works pretty well to explain the band splitting shown in the density of states plot in Fig.~\ref{fig:laddertransport}(a), and subsequently reflected in the transmission spectrum in Fig.~\ref{fig:laddertransport}(c), where we have chosen $\lambda=1$. The strength of the flux in any $n$-th plaquette is now equal to $1/2\pi$, which turns out to be a rather small perturbation , and thus the transmission pattern is not expected to change its global pattern appreciably from the zero flux case. The step-like structure , as we already have stated in the main text, exactly corroborates the results obtained analytically in case of nanoribbon strips where, for an $M$ strand ladder network one finds a total of $2M$ steps in the transmission coefficient\cite{stegmann}.

\section{The symmetry and topological edge states}
To explore any bulk symmetry and its connection to the edge states, such as the gapped states in the butterfly spectrum, shown in Fig.~\ref{fig:edge}(c), we investigate the case for $\alpha=1/3$. The existence of the other pair of edge states for $\alpha=2/3$ can be justified in a similar way.\\
\begin{figure}[h!]
	\includegraphics[width=0.9\columnwidth]{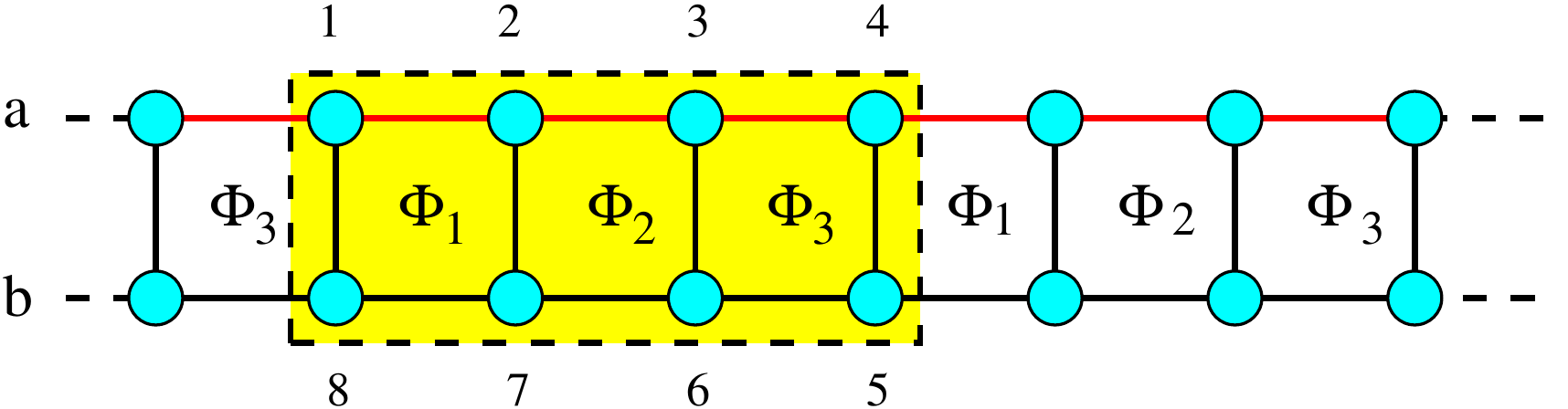}
	\caption{(Color online.) A two strand ladder threaded by a period-$3$ flux profile. The yellow shaded region represents the unit cell. Peierls' phase is acquired as the particle hops along the upper (red) bonds, and has a positive sign from left to right.}
	\label{fig:laddercell}
\end{figure}
It is to be appreciated that, the parameter $\alpha$ controls the flux modulation frequency and thus, the case $\alpha=1/3$ corresponds to a ladder with flux periodicity $N_{\phi}=\frac{1}{\alpha}=3$. This scheme of flux distribution is shown in the Fig.~\ref{fig:laddercell} with the unit cell highlighted in yellow. For $N_{\phi}=3$, the Hamiltonian in Eq.~\ref{eq:ham} can be written in the momentum space as,
\begin{widetext}
\begin{align}
\mathcal{H}(k)=
&\left[ {\begin{array}{cccccccc}
0& t_x e^{i\theta_1} & 0 & t_x e^{ik} & 0 & 0 & 0 & t_y\\
t_x e^{-i\theta_1} & 0 & t_x e^{i\theta_2} & 0 & 0 & 0 & t_y & 0 \\
0 & t_x e^{-i\theta_2} & 0 & t_x e^{i\theta_3} & 0 & t_y & 0 & 0 \\
t_x e^{-ik} & 0 & t_x e^{-i\theta_3} & 0 & t_y & 0 & 0 & 0 \\
0 & 0 & 0 & t_y & 0 & t_x & 0 & t_x e^{-ik} \\
0 & 0 & t_y & 0 & t_x & 0 & t_x & 0 \\
0 & t_y & 0 & 0 & 0 & t_x & 0 & t_x \\
t_y & 0 & 0 & 0 & t_x e^{ik} & 0 & t_x & 0 \\
\end{array} } \right]
\end{align}
\end{widetext}

We of course, have chosen $t_x=t_y=1$ throughout, but in this analysis that really does not matter. The Peierls' phase $\theta_n$ is related to the trapped magnetic flux by the relation $\theta_n= 2\pi \phi_n \quad(n=1,2,3)$,  $k$ being the wave vector. We have set the lattice translation as unity, without losing any physics. We find $\mathcal{H}(k)$ exhibits chiral symmetry, viz, $\mathcal{C}\mathcal{H}(k)\mathcal{C}^{-1}= -\mathcal{H}(k)$, with the unitary operator $\mathcal{C}$ is given by a $8 \times 8$ diagonal matrix $\mathcal{C} =$ diag $(1,-1,1,-1,1,-1,1,-1)$. Therefore, the gapped states of such bulk systems are protected by the chiral symmetry. 


\bibliography{ref.bib}

\end{document}